\documentclass[12pt]{article}
 \pdfoutput=1

\usepackage[nosort]{cite}
\usepackage{cite}
\usepackage{amsmath,pifont,bbding,graphicx}
\usepackage{epsfig}
\usepackage{amsfonts}
\usepackage{amssymb}
\usepackage{multirow}
\usepackage{enumerate}
\usepackage{subfigure}
\usepackage{slashed}

\usepackage{color}

\usepackage{subfigure}

\usepackage{caption}

\newcommand{\be}{\begin{equation}}
\newcommand{\ee}{\end{equation}}
\newcommand{\bee}{\begin{equation*}}
\newcommand{\eee}{\end{equation*}}
\newcommand{\bea}{\begin{eqnarray}}
\newcommand{\eea}{\end{eqnarray}}
\newcommand{\bean}{\begin{eqnarray*}}
\newcommand{\eean}{\end{eqnarray*}}

\begin{document}

\setcounter{page}{0}
\thispagestyle{empty}

\begin{flushright}
CERN-PH-TH/2013-028\\
UCI-HEP-TR-2013-04\\
\today
\end{flushright}

\vskip 8pt

\begin{center}
{\bf \LARGE {
Gamma Rays from
Top-Mediated \\
\vskip 8pt
Dark Matter Annihilations   }}
\end{center}

\vskip 12pt

\begin{center}
 {\bf C.~B.~Jackson$^{a}$,  G\'eraldine  Servant$^{b,c,d}$, Gabe Shaughnessy$^{e}$, \\
 \vskip 4pt
 Tim M.P. Tait$^{f}$, and Marco Taoso$^{d,g}$ }
\end{center}

\vskip 12pt

\begin{center}

\centerline{$^{a}${\it University of Texas at Arlington, Arlington, TX 76019 USA}}
\centerline{$^{b}${\it CERN Physics Department, Theory Division, CH-1211 
Geneva 23, Switzerland}}
\centerline{$^{c}${\it ICREA at IFAE, Universitat Aut\`onoma de Barcelona, 08193 Bellaterra, Barcelona, Spain}}
\centerline{$^{d}${\it Institut de Physique Th\'eorique, CEA/Saclay, F-91191 
Gif-sur-Yvette C\'edex, France}}
\centerline{$^{e}${\it Department of Physics, University of Wisconsin, Madison, WI 53706 USA}}
\centerline{$^{f}${\it Department of Physics \& Astronomy, University of California, Irvine, CA 92697 USA}}
\centerline{$^{g}${\it Department of Physics \& Astronomy, University of British Columbia,
Vancouver, BC V6T 1Z1 Canada}}
\vskip .3cm
\centerline{\tt geraldine.servant@cern.ch, chris@uta.edu,}
\centerline{\tt gshau@hep.wisc.edu, ttait@uci.edu, marco.taoso@cea.fr}
\end{center}

\vskip 10pt

\begin{abstract}
\vskip 3pt
\noindent

Lines in the energy spectrum of gamma rays are a fascinating experimental signal,
which are often considered ``smoking gun" evidence of dark matter annihilation.
The current generation of gamma ray observatories are currently closing in on
parameter space of great interest in the context of dark matter which is a thermal
relic.  
We consider theories in which the dark matter's primary connection
to the Standard Model is via the top quark, realizing strong gamma ray lines
consistent with a thermal relic through the forbidden channel mechanism
proposed in the Higgs in Space Model.
We consider realistic UV-completions of the Higgs in Space and related
theories, and show that a rich structure of observable gamma ray lines
is consistent with a thermal relic as well as constraints from dark matter
searches and the LHC.   
Particular attention is paid to the
one loop contributions to the continuum gamma rays, which can easily
swamp the line signals in some cases, and have been largely overlooked
in previous literature.

\end{abstract}

\newpage


\vskip 13pt

\section{Introduction}

The existence of nonbaryonic dark matter is well established, but there is still no clue as to its particle identity, and no
indication as to what interactions beyond gravitational it experiences.  In order to place dark matter in context within
the Standard Model or its extensions, observation of dark matter interacting through something other than gravity will
be required.  There is currently a rich and diverse experimental program searching for dark matter through its
direct scattering with nuclei, produced at colliders, or indirectly through its annihilation products.

This last pillar of the searches for dark matter is particularly interesting.  Observation of dark matter annihilation would
be a clear indication that dark matter has non-gravitational interactions, and (depending on the observations) would help
to establish it as a thermal relic.  In fact, observations of dwarf spheroidals by the Fermi LAT currently
exclude $s$-wave annihilating
thermal relic dark matter with mass below about 50~GeV~\cite{Ackermann:2011wa,Ackermann:2012qk}.

An important element in the indirect search for dark matter is the search for lines in the energy spectrum of gamma rays.
Such features occur when dark matter annihilates into a two-particle final state, with one of the particles a photon.  Since
dark matter must be electrically neutral (to good approximation \cite{McDermott:2010pa}), such a signal is typically
expected to be rather smaller than the expected annihilation cross section of a thermal relic,
 $\langle \sigma v \rangle \sim 3 \times 10^{-26}$~cm$^2$/s (with mild dependence on the dark matter
 mass \cite{Steigman:2012nb}).  Nonetheless, such a gamma ray line is a very distinctive
 signal, unlikely to be faked by astrophysical backgrounds. In preparation for further data from Fermi, as well
 as new data sets from H.E.S.S., CTA, and other future gamma ray observatories,
 it is worthwhile to explore theories which would be expected to
 result in a prominent line signal.
 
Typically, there is some tension between the need to have a large loop level line signal without ending up with a
large annihilation cross section into continuum photons, which could either swamp the line signal, or would over-saturate
the thermal cross section.  An early step in this direction was the ``Higgs in Space" model \cite{Jackson:2009kg},
which was inspired by models of dark matter in the setting of warped extra dimensional GUTs  
\cite{Agashe:2004ci,Agashe:2004bm}, and had a Dirac fermion dark matter which annihilated through a $Z^\prime$
which coupled preferentially to top quarks.  By annihilating into top quarks, one could arrange for a thermal relic
with masses slightly below the top mass which allowed for efficient annihilation in the early universe, but suppressed
continuum annihilation in cold structures such as galaxies today.

Ingredients such as $s$-channel resonant annihilation and forbidden channels in the final state
have received much attention recently  \cite{Buckley:2012ws}, largely in response to the reported excess of gamma rays
around $\sim 130$~GeV originating close to the galactic center in the Fermi LAT data
\cite{Weniger:2012tx,Tempel:2012ey,Su:2012ft,Rao:2012fh}.  While interest was inflamed by a subdominant signal at
around 115~GeV consistent with an additional annihilation channel into $\gamma Z$
\cite{Su:2012ft,Rajaraman:2012db,Cohen:2012me}, peculiar features
such as an excess of photons at the same energy from the Earth's limb
\cite{Su:2012ft,Finkbeiner:2012ez,Hektor:2012ev} could point to an instrumental origin for the feature.
While many of the constructions we consider could potentially explain the Fermi signal should it turn out to
actually be dark matter, we remain agnostic about its origin and consider theories with enhanced lines in general.

In a previous paper \cite{Jackson:2013pjq}, we considered the forbidden channel mechanism in the context of
dark matter which annihilates through exotic states with only weak connection to the Standard Model itself.  We made
the important observation that for this class of models, the one-loop continuum annihilation channels cannot
typically be ignored, and often play an important role in determining both the observability of a line signal 
as well as the relic density.
In the current
article, we revisit models closer in spirit to the Higgs in Space, where the role of the SM-charged mediator is 
once again played by the top quark.  We consider UV-complete models, and derive predictions for gamma ray lines,
as well as conditions for a thermal relic, constraints from direct detection, precision measurements, and the LHC.

Our article is outlined as follows: in Section \ref{sec:topmodel} we set up the basic module containing the
necessary ingredients, building on the experience of Ref.~\cite{Jackson:2013pjq}.
In Section~\ref{sec:precision}, we compute the constraints from precision measurements and the observed
properties of the newly
discovered Higgs boson, which help narrow down the viable parameter space,
as well as direct constraints from LHC searches.
In Section~\ref{sec:annihilation}, we compute the rates for dark matter annihilation at one loop,
including contributions to the continuum gammas, gamma ray lines, and the relic density.
Finally, we conclude in Section~\ref{sec:conclusions}.  The appendices contain detailed expressions
for effective vertices at one loop, complementing those already presented in Ref.~\cite{Jackson:2013pjq}.

\section{Top as a Messenger to Dark Matter}
\label{sec:topmodel}

The basic set-up builds on the models explored in ~\cite{Jackson:2013pjq} : the dark matter
is a Dirac\footnote{One could engineer Majorana dark matter, by involving another fermion charged under $U(1)'$, but this would require $\nu$-$Z'$ axial interactions,
and would generally lead to a large annihilation into $gg$ \cite{Jackson:2013pjq}.}
fermion $\nu$ with no SM gauge interactions, but charged under a $U(1)^\prime$
symmetry\footnote{We consider variations where the dark matter $\nu$ either has vector-like $U(1)^\prime$
interactions, in which case one can write down a gauge-invariant mass, or have chiral interactions, in which case
the mass for the dark matter will have to be generated by the $U(1)^\prime$-breaking VEV.}. 
The SM (with the subtle exception of the top quark as outlined below) is uncharged under the
$U(1)^\prime$, resulting in very weak bounds from precision measurements \cite{Carena:2004xs}. 
The $U(1)^\prime$ is higgsed by a scalar particle, resulting in a massive $Z^\prime$
vector boson and a Higgs-like state $\Phi$ (which is not to be confused the SM Higgs boson $h$).
For simplicity, we assume that any kinetic mixing between the $U(1)^\prime$ and the SM hypercharge
interaction is small enough to be ignored.
Both the $Z^\prime$ and $\Phi$ can act as mediators between the dark matter and some additional
fermions $\psi$ which are charged under the Standard Model.
In Ref.~\cite{Jackson:2013pjq}, we restricted our attention to the case where $\psi$ was electroweakly-
but not color-charged; this avoided constraints from the LHC and prevented
annihilation into gluons.  In this article, we consider the case where the role of $\psi$ is played
by an extended top sector.

We denote by $\hat{t}_R$ the field which will largely become the right-handed top quark,
uncharged under the $U(1)^\prime$ and in the $(3, 1, 2/3)$ representation under the SM
($SU(3)$, $SU(2)$, $U(1)$) symmetries.  Similarly, $\hat{Q}_3$ is the usual left-handed
quark doublet for the third generation.  The additional fermions $\psi_{L,R}$ have
equal charges under $U(1)^\prime$ and are charged as $(3, 1, 2/3)$ under the SM.  By introducing
a vector-like pair of $\psi$ fields, we avoid disrupting the cancellation of any of the purely
SM anomalies.  The $\psi$ fields are also vector-like under $U(1)^\prime$, and thus no mixed
SM-$U(1)^\prime$ anomalies are generated.  The last remaining potential gauge anomalies 
(depending on the charges of $\nu_L$ and $\nu_R$) are
$U(1)^\prime$-gravity$^2$ or $U(1)^{\prime 3}$, and can be arranged by including additional SM singlet fermions
which are charged under $U(1)^\prime$.  When present, these fermions are essentially irrelevant for the phenomena
of interest to us, and we will ignore them in the subsequent discussion.

\begin{figure}[t!]
\begin{center}
\includegraphics[angle=0,width=0.79\linewidth]{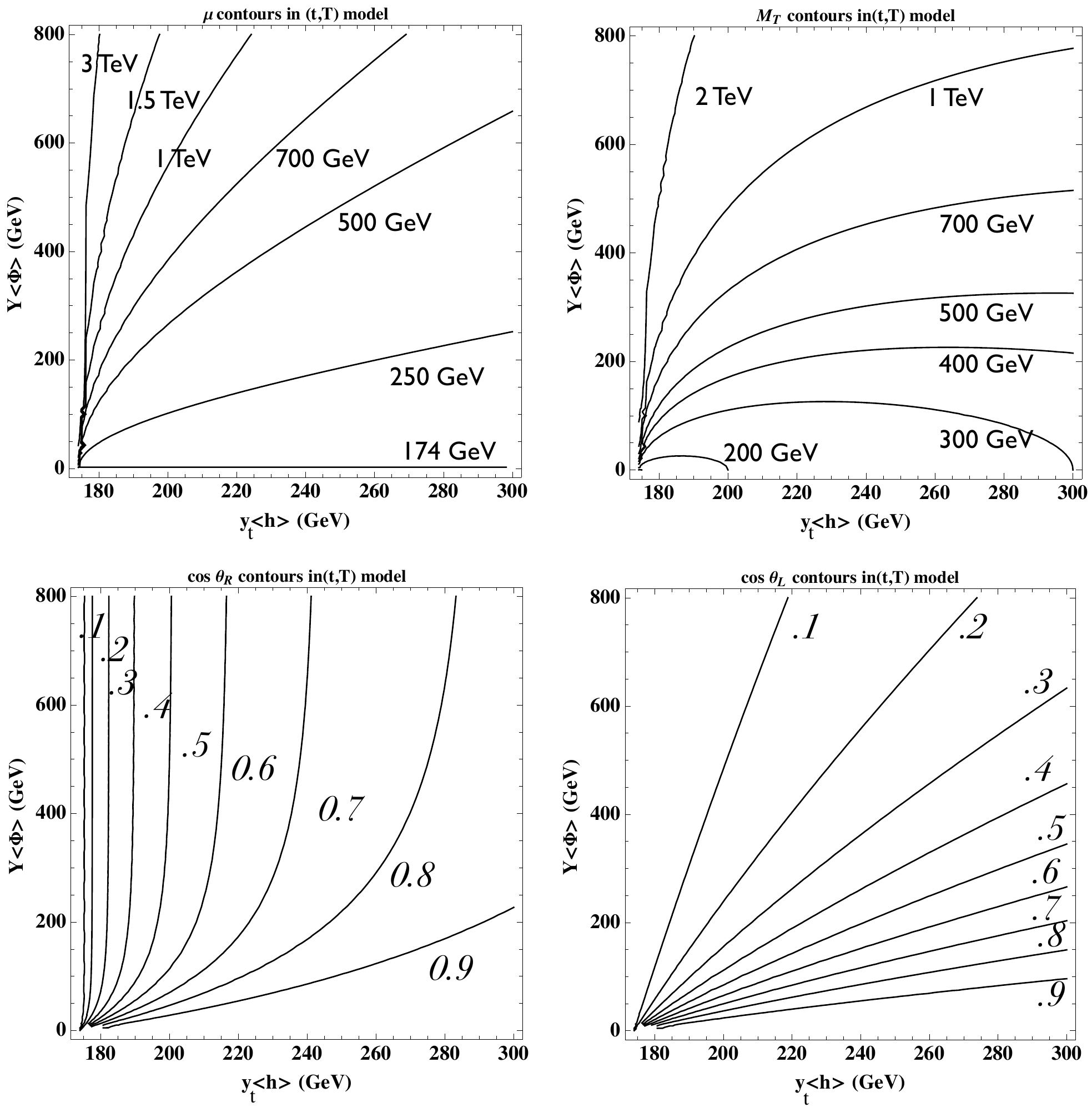}
\caption{ Contours of $\mu$, $M_T$, $\cos \theta_R$ and $\cos \theta_L$ in the top partner model.
\label{fig:contours} }
\end{center}
\end{figure}

The terms responsible for top-$\psi$ mixing are structurally reminiscent of top-seesaw models
\cite{Dobrescu:1997nm,Chivukula:1998wd,He:1999vp,He:2001fz},
\bea
{\cal L}_{\rm mass} & = & 
y H \bar{\hat{Q}}_3 \hat{t}_R  + \mu \bar{\psi}_L \psi_R + Y \Phi \bar{\psi}_L \hat{t}_R
\label{eq:lagrangian}
\eea
where $H$ is the SM Higgs doublet, $y$ and $Y$ are dimensionless couplings,
and $\mu$ is a gauge-invariant mass term for $\psi$.
We define mass eigenstates $t$ and $T$:
 \begin{equation}
 \left( \begin{array}{c}
t_{R/L}   \\
T_{R/L}    \end{array} \right)
=
  \left( \begin{array}{cc}
-\sin \theta_{R/L} & \cos \theta_{R/L}  \\
\cos \theta_{R/L} & \sin \theta_{R/L}  \end{array} \right)
\left( \begin{array}{c}
\hat{t}_{R/L}   \\
\Psi_{R/L} \end{array} \right) ~,
\label{eq:matrix}
\end{equation}
with eigenmasses,
 \begin{equation}
 M^2_{t/T}=\frac{1}{2}\left[  \mu^2+y^2\langle H^2\rangle  +Y^2\langle \Phi^2\rangle \mp \sqrt{-4 \mu^2
 y^2\langle H^2\rangle+( \mu^2+y^2\langle H^2\rangle  +Y^2\langle \Phi^2\rangle  )^2 }       \right]~,
 \end{equation}
and mixing angles $\theta_R$ and $\theta_L$ given by,
 \begin{eqnarray}
 \nonumber
 \tan \theta_R&=& \frac{\mu^2-y^2\langle H^2\rangle  -Y^2\langle \Phi^2\rangle + \sqrt{-4 \mu^2
 y^2\langle H^2\rangle+( \mu^2+y^2\langle H^2\rangle  +Y^2\langle \Phi^2\rangle  )^2 }  }{2 \mu Y \langle \Phi \rangle}\\
 \tan \theta_L&=& \frac{\mu^2-y^2\langle H^2\rangle  +Y^2\langle \Phi^2\rangle + \sqrt{-4 \mu^2
 y^2\langle H^2\rangle+( \mu^2+y^2\langle H^2\rangle  +Y^2\langle \Phi^2\rangle  )^2 }  }{2 y \langle H \rangle Y \langle \Phi \rangle}~.
 \label{eq:tan}
 \end{eqnarray}
We identify the lighter mass eigenstate with the top quark of mass $m_t \simeq 174$~GeV, 
leaving two free parameters which we choose to be $y\langle H \rangle$ and $Y \langle \Phi \rangle$.
Imposing the top mass constraint, $\mu$ is given by,
 \begin{equation}
 \mu^2=m_t^2\frac{m_t^2-y^2\langle H^2\rangle -Y^2\langle \Phi^2\rangle }{m_t^2-y^2\langle H^2\rangle}
 \end{equation}
 and $M_T$ is 
 determined
We show the contours of $\mu$, $M_T$, $\cos \theta_R$ and $\cos \theta_L$ in Fig.~\ref{fig:contours}.
It is worth noting that in the limit $Y\langle  \Phi \rangle  \to 0$, 
the mixing angles do not go to zero (unless one also requires $y \langle H \rangle  \to m_t$). 
For a fixed $y \langle H \rangle \neq m_t$,  the limit
$Y \langle \Phi \rangle \to 0$ necessarily implies $\mu\to m_t $ in order
to keep the lightest mass eigenstate at $m_t =174$ GeV.  In addition, in the SM-like
regime where $c_R, c_L \ll 1$ and the second mass eigenstate is very heavy is obtained for
$Y \langle \Phi \rangle, \mu \to \infty$. 
Finally,
the case where $y \langle H \rangle \sim m_t$ and $M_T \gg m_t$ corresponds to a value of 
$Y \langle \Phi\rangle $ which is parametrically smaller than $M_T$ itself.

In the mass basis one has diagonal interactions of $t$ and $T$ as well as mixed $t$-$T$ interactions,
 \begin{eqnarray}
 {\cal L}&=& i \ \bar{t}\slashed{D}_t t + i \  \bar{T}\slashed{D}_T T 
 + \frac{g}{\sqrt{2}} (c_L \bar{T} - s_L\bar{t}) \gamma^{\mu} W_{\mu} P_L  \ b       + h.c \\
 \nonumber
& +& s_L c_L [g_{\hat{t}_R}^Z-g_{\hat{t}_L}^Z]  \ \bar{T} \gamma^{\mu} Z_{\mu}  \ P_L  \ t + h.c \\
\nonumber
& +& \bar{T} \gamma^{\mu}  \hat{Z}^{\prime}_{\mu} \ g_{\psi}^{\hat{Z}^{\prime}} (s_L c_L  P_L + s_R c_R  P_R  ) \ t + h.c \\
\nonumber
& + & y s_L s_R \  h \ \bar{{t}}  \ {t}  + y c_L c_R \  h \ \bar{{T}}  \ {T} - y \ h \  \bar{t}(c_Rs_L P_R+c_Ls_R P_L) T + h.c\\
\nonumber
& - & Y c_L s_R \  \varphi \ \bar{{t}}  \ {t}  + Y s_L c_R \  \varphi \ \bar{{T}}  \ {T} + Y \ \varphi \  \bar{t}(c_Lc_R P_R-s_Ls_R P_L) T + h.c
  \end{eqnarray}
 where
 \begin{equation}
{D_t}_{\mu}=\partial_{\mu}  - i \left[ (g^Z_{\hat{t}_R} c_L^2 + g^Z_{\hat{t}_L} s_L^2)P_L+g^Z_{\hat{t}_R}P_R
\right] Z_{\mu} 
-i g_{\psi}^{\hat{Z}'}(c_R^2P_R+c_L^2P_L) \hat{Z}^{\prime}_{\mu}
-i g_{\hat{t}_R}^{\gamma}A_{\mu}
  \end{equation}
  and,
  \begin{equation}
{D_T}_{\mu}=\partial_{\mu} - i \left[ (g^Z_{\hat{t}_R} s_L^2 + g^Z_{\hat{t}_L} c_L^2)P_L+g^Z_{\hat{t}_R}P_R
\right] Z_{\mu} 
-i g_{\psi}^{\hat{Z}'}(s_R^2P_R+s_L^2P_L) \hat{Z}^{\prime}_{\mu}\\
-i g_{\hat{t}_R}^{\gamma}A_{\mu}
  \end{equation}
with  $g_{\hat{t}_R}^{\gamma}= 2e/3$,  $g_{\hat{t}_{R/L}}^{Z}=e \cos \theta_W^{-1} \sin \theta_W^{-1} 
(T_3- (2e/3)\sin^2\theta_W)$  and $g=e/\sin \theta_W$ the usual gauge couplings.  In addition, both
$t$ and $T$ have the usual diagonal color-triplet couplings to gluons, as enforced by $SU(3)$ gauge invariance.
We have introduced the short-hand notation $c_L \equiv \cos \theta_L$, $s_R \equiv \sin \theta_R$, etc.
  
 \section{Precision and LHC Constraints}
 \label{sec:precision}
 
 Before turning to the annihilation cross sections and resulting line signals, we examine the constraints on the parameter
 space from precision measurements, LHC searches and direct DM experiments.

\subsection{Oblique Corrections}
 
 \begin{figure}[t]
\begin{center}
\includegraphics[width=0.495\textwidth]{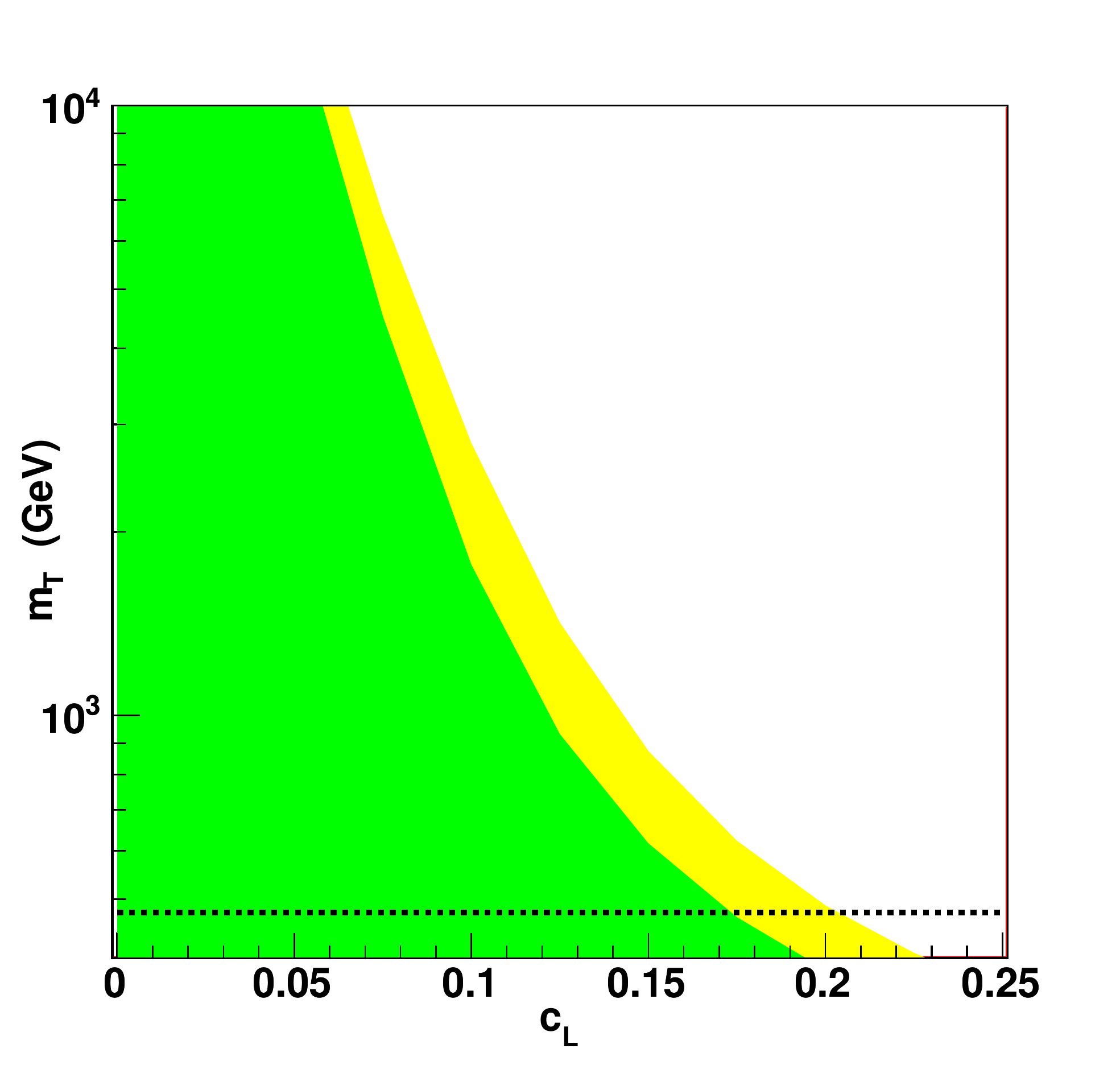}
\includegraphics[width=0.495\textwidth]{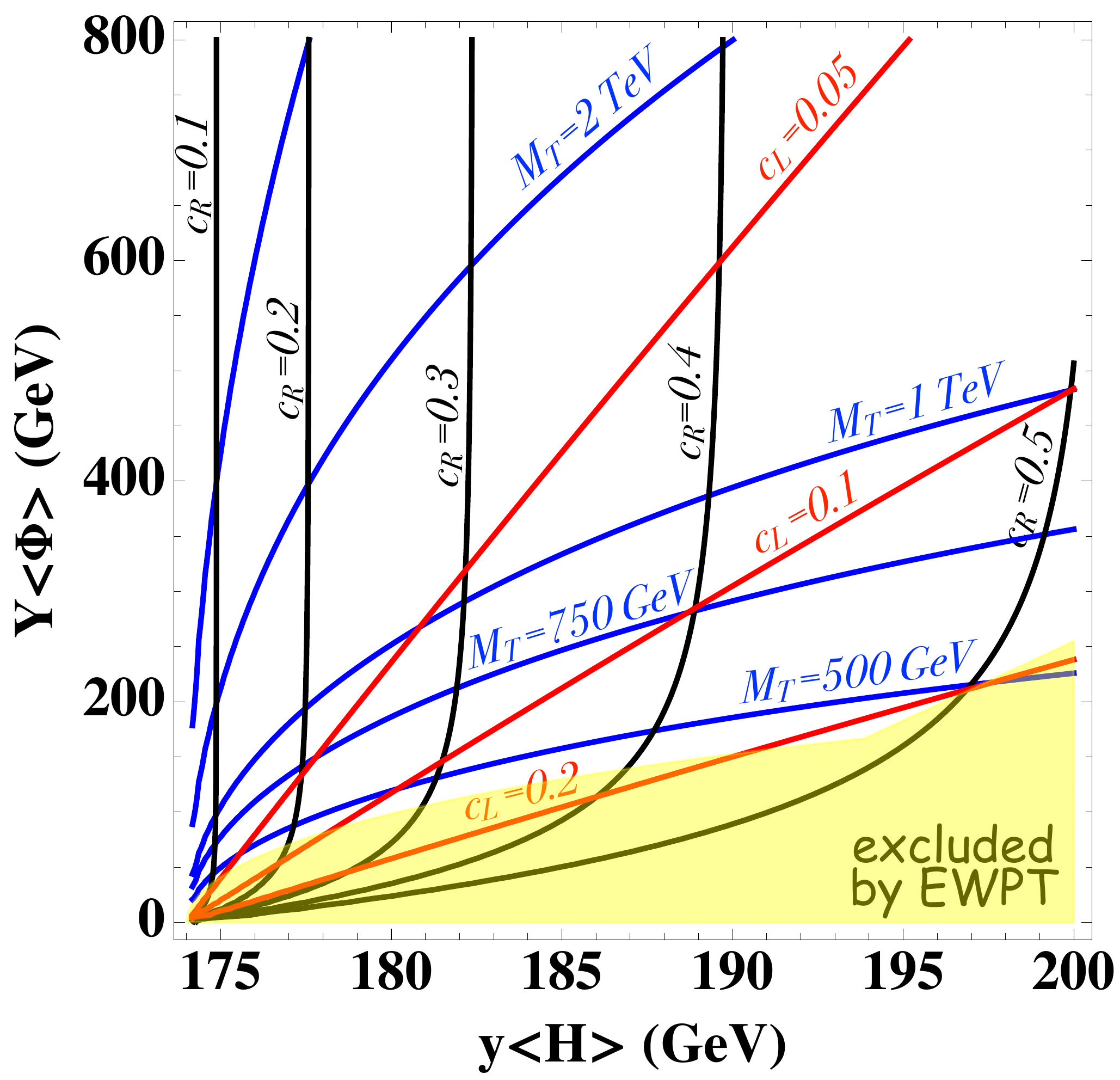}
\caption{(a) Region of the parameter space of $M_T$ and $c_L$ consistent with precision
electroweak data at 1$\sigma$ (green region) and $95\%$ confidence level (yellow region).
Also shown is the approximate lower bound on $M_T$ from null LHC searches (dashed line); (b) Contour 
plots for $c_L, c_R$ and  $M_T$ in the $(\langle y H\rangle, \langle Y \Phi\rangle)$ plane where the yellow region is  excluded by EW data. }
\label{fig:ewconstraints}
\end{center}
\end{figure}

 As a consequence of mixing with $T$, 
 the $Z$-$\bar{t_L}$-$t_L$ and  $W$-$\bar{t_L}$-$b_L$ EW interactions of the top quark
 are modified with respect to the Standard Model; these deviations are small in the unmixed limit of $c_L \rightarrow 0$. 
 Because of the modified EW interactions of top, as well as diagrams containing the partner quark $T$ explicitly,
 there are oblique contributions to the precision electroweak observables encapsulated by the
 Peskin-Takeuchi parameters \cite{Peskin:1991sw} which provide bounds on the mixing parameter $c_L$ and
 $M_T$.  The relevant corrections were computed in Refs.~\cite{He:2001fz,Dawson:2012di}, where it was found
 that $\Delta S$ and $\Delta U$ are very small, and the dominant constraint is from 
$\Delta T$,
\bea
\Delta T & = & T_{SM} \times
c_L^2 \left( -(1 + s_L^2) + c_L^2 r + 2 s_L^2 \frac{r}{(r - 1)} \log r \right)
\eea
where,
\be
r  \equiv  \frac{m_T^2}{m_t^2}, \ \mbox{and} \ \ 
T_{SM} = \frac{3}{16 \pi s_W^2} \frac{m_t^2}{M_W^2} \simeq 1.19~.
\ee
For a Higgs mass of $m_h = 125.7 \pm 0.4$~GeV, a global fit to the electroweak
data restricts $\Delta T \leq 0.14~(0.10)$ for $\Delta S \simeq \Delta U \simeq 0$ 
at the $95\%~(68\%)$ confidence
level \cite{Baak:2012kk}.  For a fixed value of the mixing $c_L$, this corresponds to an upper bound on
$M_T$.  The constrained region is plotted in Fig.~\ref{fig:ewconstraints}(a), and indicates that for
$c_L \sim 0.1$, masses $M_T \leq 3$~TeV are consistent with precision measurements.
Also shown on the figure is the line corresponding to $M_T \geq 475$~GeV, which is a conservative
estimate for the lower bound on the mass of the $T$ quark based on a collider searches (see below).

We show in Fig.~\ref{fig:ewconstraints}(b) the contour plots for $c_L, c_R$ and  $M_T$  in the 
parameter space of $y \langle H \rangle$ and $Y \langle \Phi \rangle$, with the region excluded by
precision measurements overlaid.
In the allowed region, $c_L$ is smaller than $\sim 0.2$, but 
$c_R$ may still be sizable provided $T$ is not too heavy.
 
\subsection{Constraints from Higgs Measurements} 

\begin{figure}[t!]
\begin{center}
\includegraphics[angle=0,width=0.99\linewidth]{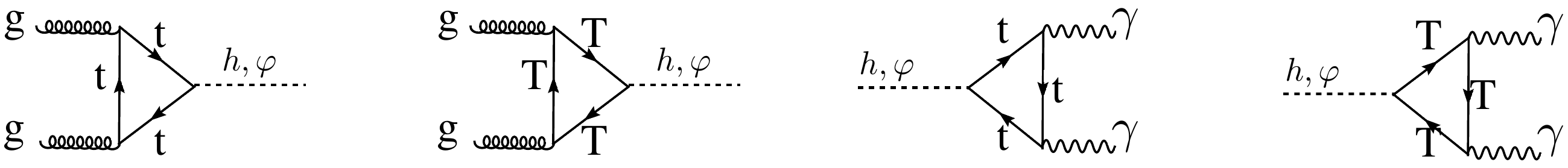}
\end{center}
\caption{$h$ and $\varphi$  production by gluon fusion and their decay into $\gamma \gamma$.}
\label{fig:Higgs} 
\end{figure}

\begin{figure}[t!]
\begin{center}
\includegraphics[width=0.4\textwidth]{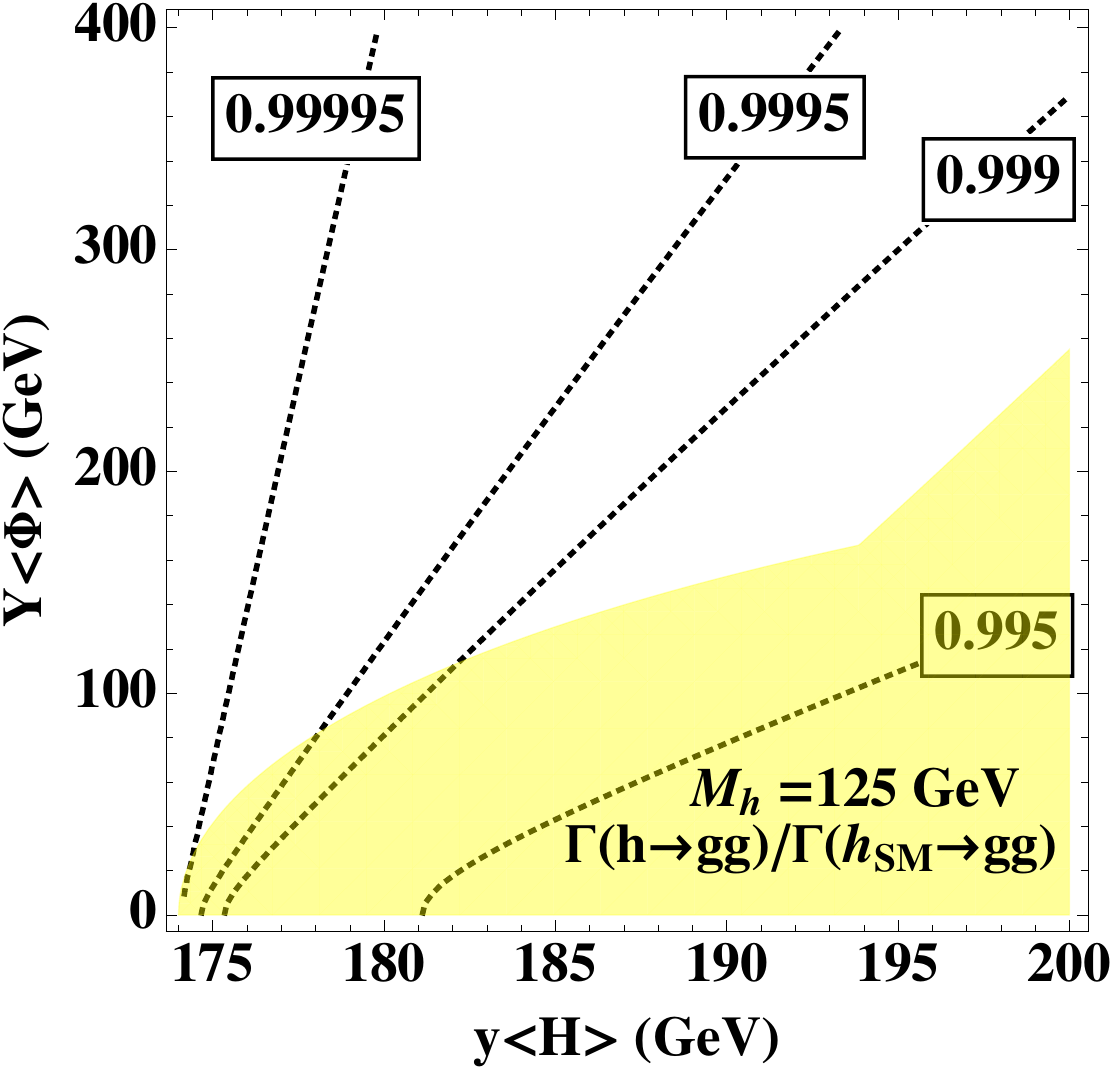}
\caption{ Contours for $\Gamma_{h\to gg}/\Gamma^{SM}_{h \to gg}$. The yellow region is excluded by
precision electroweak measurements.}
\label{fig:hprod}
\end{center}
\end{figure}

In addition, there can be constraints on $c_L$ and $c_R$ from production of the newly
discovered Higgs boson \cite{:2012gk}
at the LHC, which receives
modifications due to the shift in the top coupling to the Higgs, as well as corrections from the $T$ quark itself,
which contributes to Higgs interactions with gluons and photons (see Fig.~\ref{fig:Higgs}).  However, in the region
consistent with precision electroweak constraints,
Higgs phenomenology remains almost very close to 
Standard Model-like at the LHC.  For example, the corrections to $h \rightarrow g g$ (which also characterize the
shift in inclusive Higgs production) are,
\be
\Gamma_{h\to gg} \propto \left| \frac{s_L s_R F_{1/2}(\tau_t)}{m_t/v}
+\frac{c_L c_R F_{1/2}(\tau_T)}{m_T/v} \right|^2 \propto \left| \frac{F_{1/2}(\tau_t)}{m_t/v} \right|^2~,
\ee
where the function $F_{1/2}$ of $\tau_q \equiv m_h^2 / (4 m_q^2)$ can be found in Ref.~\cite{Djouadi:1991tka}.
As illustrated in Fig.~~\ref{fig:hprod}, these are much less than $1\%$ deviations for the parameter space of interest.  Such
tiny deviations might be measurable at a future linear collider and/or Higgs factory, but are unlikely to be accessible
at the LHC.  Similarly, the correction to $h \rightarrow \gamma \gamma$ is likely to be even smaller since it is
dominantly mediated via a $W$ loop, and would
be extremely difficult to detect.

One could also search for the $\Phi$ boson, which, once produced, would decay via its mixing with the Higgs.
However, the $t$ and $T$ loops contributing to $\Phi$ production at the LHC  almost 
cancel since $\tan \theta_R/\tan \theta_L \equiv  m_t/m_T$, and
\be
\Gamma_{\varphi \to gg} \propto \left| -\frac{c_L s_R F_{1/2}(\tau_t)}{m_t/v}+\frac{s_L c_R F_{1/2}(\tau_T)}{m_T/v} \right|^2\propto 
\frac{ c_L^2 c_R s_R}{m_t} |F_{1/2}(\tau_t)|^2 \left| 1-\frac{F_{1/2}(\tau_T)}{F_{1/2}(\tau_t)} \right|^2,
\ee
resulting in the production rate of $\Phi$ being very suppressed.

\subsection{Constraints from LHC Searches}
\label{sec:LHC}

\begin{figure}[t]
\begin{center}
\includegraphics[width=0.45\textwidth]{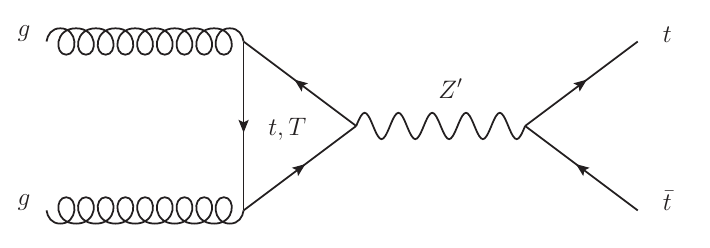}\\
\includegraphics[width=0.35\textwidth]{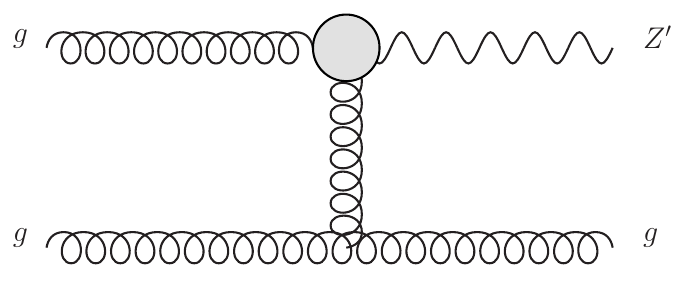}
\includegraphics[width=0.35\textwidth]{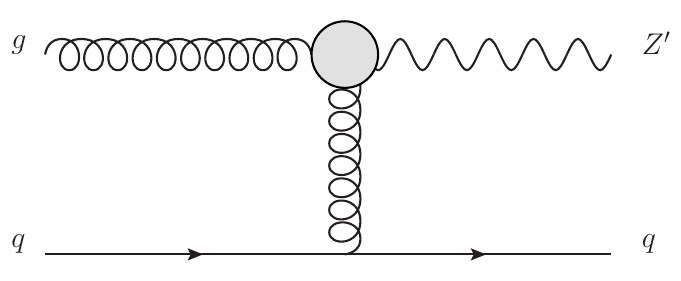}
\caption{Production diagrams for $Z^\prime$ through its induced coupling to gluons.}
\label{fig:ttbarprod}
\end{center}
\end{figure}

The primary constraints from the LHC are from the null searches for exotic quarks such as our $T$.  As a vector-like
quark which mixes with the top, the primary decays are expected to be: $T \rightarrow W b$, $T \rightarrow Z t$, and
$T \rightarrow h t$.  In addition, if either $\Phi$ or $Z^\prime$ are light enough, one could have decays into either one
of them plus a top quark.  Null searches for the $Wb$, $Zt$, and $ht$ channels at the LHC result in a conservative
(in the sense that larger masses are permitted for any configuration of branching ratios)
bound of $M_T \geq 475$~GeV \cite{Rao:2012gf,CMS:2012ab}.
In the case where $T$ decays entirely into $t h$, the current bound derived from ATLAS data is
425 GeV \cite{Rao:2012gf}.

In addition, there are constraints on various types of $Z^\prime$ bosons that may be inferred from LHC searches.
The primary interaction of the $Z^\prime$ with the Standard Model is with top quarks.  As we shall see, the result
is suppressed production cross sections which are generically safe from LHC bounds.

Given the large coupling to top quarks, the dominant $Z^\prime$ production is via gluon fusion through a top
loop, as in Figure~\ref{fig:ttbarprod}.  
Detailed expressions for the effective $Z^\prime$-$g$-$g$ vertex 
are provided in Appendix \ref{sec:appendixttbar}.
The $Z^\prime$ can subsequently decay into $\nu \bar{\nu}$ leading
to a missing energy signature, or back into top quarks leading to a resonance in the invariant mass distribution
of $t \bar{t}$ pairs.  
An interesting feature of this signal arises because the Landau-Yang
theorem \cite{Yang:1950rg}
insures that the amplitude must vanish when the $Z^\prime$ is on-shell.  As a result, the signal is much smaller than
one might naively expect, and would have an interesting shape around the $Z^\prime$ resonance if it were
visible and the $Z^\prime$ width were large enough to permit it to be experimentally mapped out.  However,
from the point of view of constraining the parameters, 
one can derive more stringent bounds by considering the process where one of the initial gluons
is highly off-shell, resulting in a final state consisting of the $Z^\prime$ and an additional hadronic jet
(and after the $Z^\prime$ decays to $t \bar{t}$ leads to a $t \bar{t} j$ signature).

The choice of $p_T$ cut on the associated jet can be optimized based on the need to pull one of the initial gluons
off-shell so that the $Z^\prime$ production 
rate escapes from the Landau-Yang suppression, while at the same time not choosing a cut so stringent as
to result in a large suppression from the parton distribution functions.  Fully optimizing the search cuts is beyond the
scope of this work, but for a choice of $p_T > 30$~GeV, large enough to suppress the $t \bar{t}$ background and
such that collinear logarithms do not spoil perturbation theory, we find that even for large couplings, the
deviation at $\sqrt{s} = 8$~TeV
is never more than about $0.1\%$ of the SM $t \bar{t} j$ rate\footnote{When 
the $Z^\prime$ can decay both into $\nu\bar{\nu}$ and $t\bar{t}$,
the branching fraction for $Z^\prime \to t\bar{t}$ is less than $10^{-2}$, since the top
coupling is suppressed by $c_R^2$ compared to the  $\nu$ coupling.}, with the largest potential deviations occurring
when the the $Z^\prime$ mass is just above the $t \bar{t}$ decay threshold.  For larger values of the $p_T$ cut,
the deviation is even smaller.  Such deviations are well within the uncertainties in the SM rate from theoretical
uncertainties such as the gluon parton flux and scale variation, and thus will be very difficult to extract from the
background.  This situation is essentially unchanged at $\sqrt{s} = 14$~TeV.

When the $Z^\prime$ decays directly into dark matter, it produces a dark matter mono-jet 
signature \cite{Goodman:2010yf}.
For very heavy $Z^\prime$s, well-represented by a contact interaction, the current LHC bounds are a factor of few too 
weak to bound our construction \cite{ATLAS:2012ky}, but future
prospects in this channel would appear to be good.
As the $Z^\prime$ gets lighter and is no longer well-approximated by a contact interaction, 
these bounds are typically somewhat relaxed \cite{Bai:2010hh}.

The $Z^\prime$ could also be produced through its induced electroweak interactions, through
processes such as $q \bar{q} \rightarrow V^* \rightarrow Z^\prime V^\prime$
(where $V,V^\prime = \gamma, Z$), which was
considered in Ref.~\cite{Lee:2012ph} in the context of the Fermi 130 GeV line feature.  For
$Z'$-$Z$-$\gamma$  couplings large enough to explain the Fermi feature, the LHC can provide useful bounds (though
it is somewhat questionable whether such large effective interactions are consistent with
a perturbative UV complete theory) in the regime of weak DM-$Z'$ couplings, but in our construction the effective vertex is small enough
that there is essentially no useful bound from this channel, even for very light $Z^\prime$s.

A $Z^\prime$ with large coupling to top quarks can also be produced as radiation from a $t \bar{t}$ pair through processes
such as $g g \rightarrow t \bar{t} Z^\prime$.  Its subsequent decay into dark matter leads to a top pair plus missing
energy signature \cite{Cheung:2010zf}, or it can decay back into top quarks leading to a four top final state 
\cite{Bevilacqua:2012em,Lillie:2007hd,Pomarol:2008bh,Servant:2010zza}.  
For either process, the rate is modest because of the phase space suppression and 
is unlikely to be a serious constraint from current LHC data.  However, prospects are good for observation at higher
energy LHC running.

\subsection{Constraints from Direct Detection}
\label{sec:direct}

For low $Z^\prime$ masses, constraints from null results of
direct searches for dark matter (e.g. from Xenon-100 \cite{Aprile:2012nq}) 
can be potentially rather strong.  
In particular, it implies a bound on the degree of kinetic
mixing between the $Z$ and $Z^\prime$, since that parameter controls the coupling
to light quarks\footnote{There is also direct scattering through the induced coupling to gluons
mentioned above, but it leads to predictions well below the current bounds.}.
As the direct scattering of $\nu$ with nuclei is entirely on protons
rather than neutrons, the Xenon-100 bounds (which assume equal scattering
with both protons and neutrons) require $\sigma_{p\nu} \lesssim 3 \times 10^{-44}$ cm$^2$,
corresponding to the restriction $\eta \times g_{Z'}\lesssim 10^{-3}$ \cite{Jackson:2009kg},
where $\eta$ is the coefficient of the $F_Y^{\mu \nu} F^\prime_{\mu \nu}$ term in the effective action.

\begin{figure}[t!]
\begin{center}
\includegraphics[angle=0,width=0.45\linewidth]{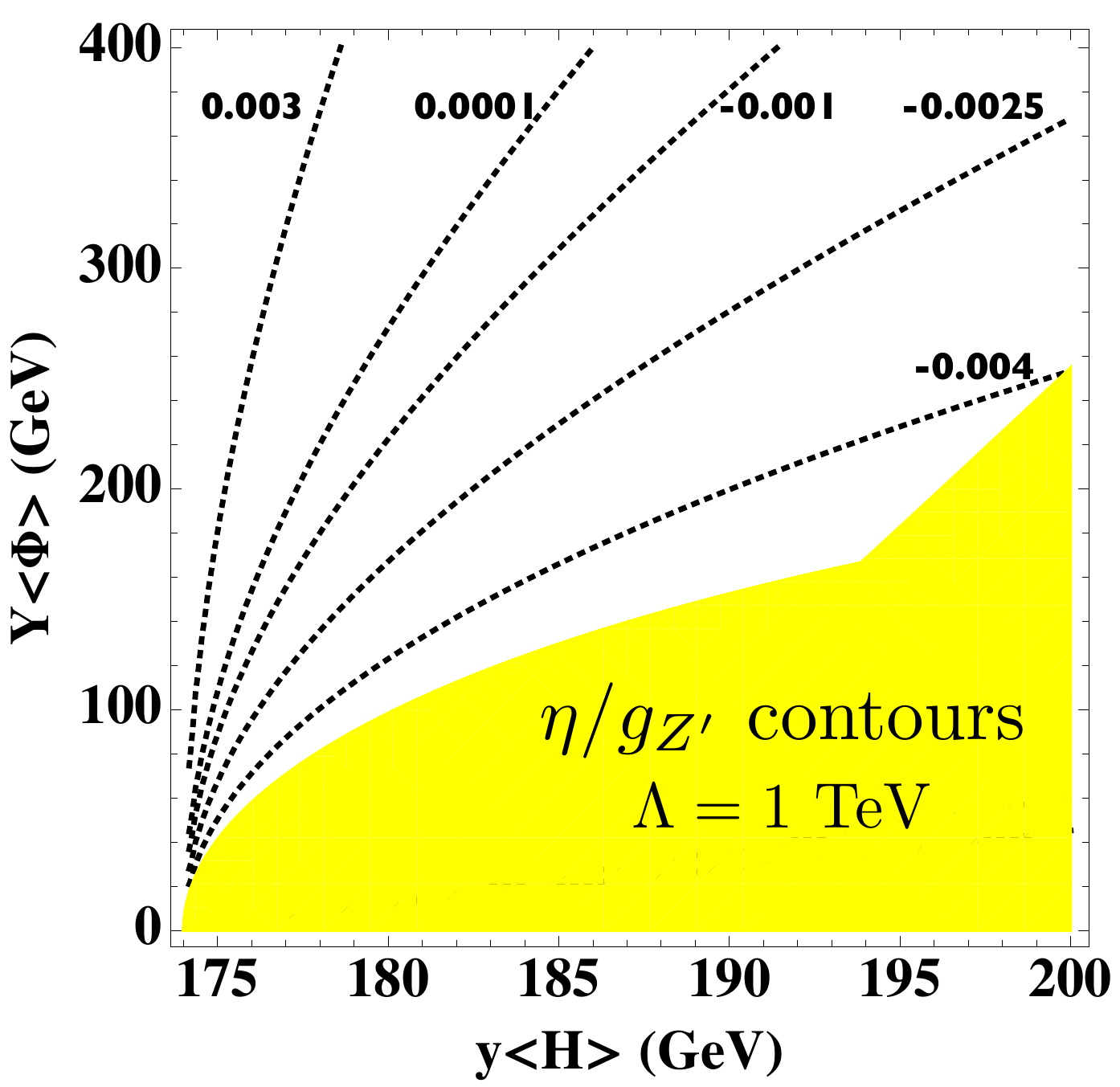}
\includegraphics[angle=0,width=0.45\linewidth]{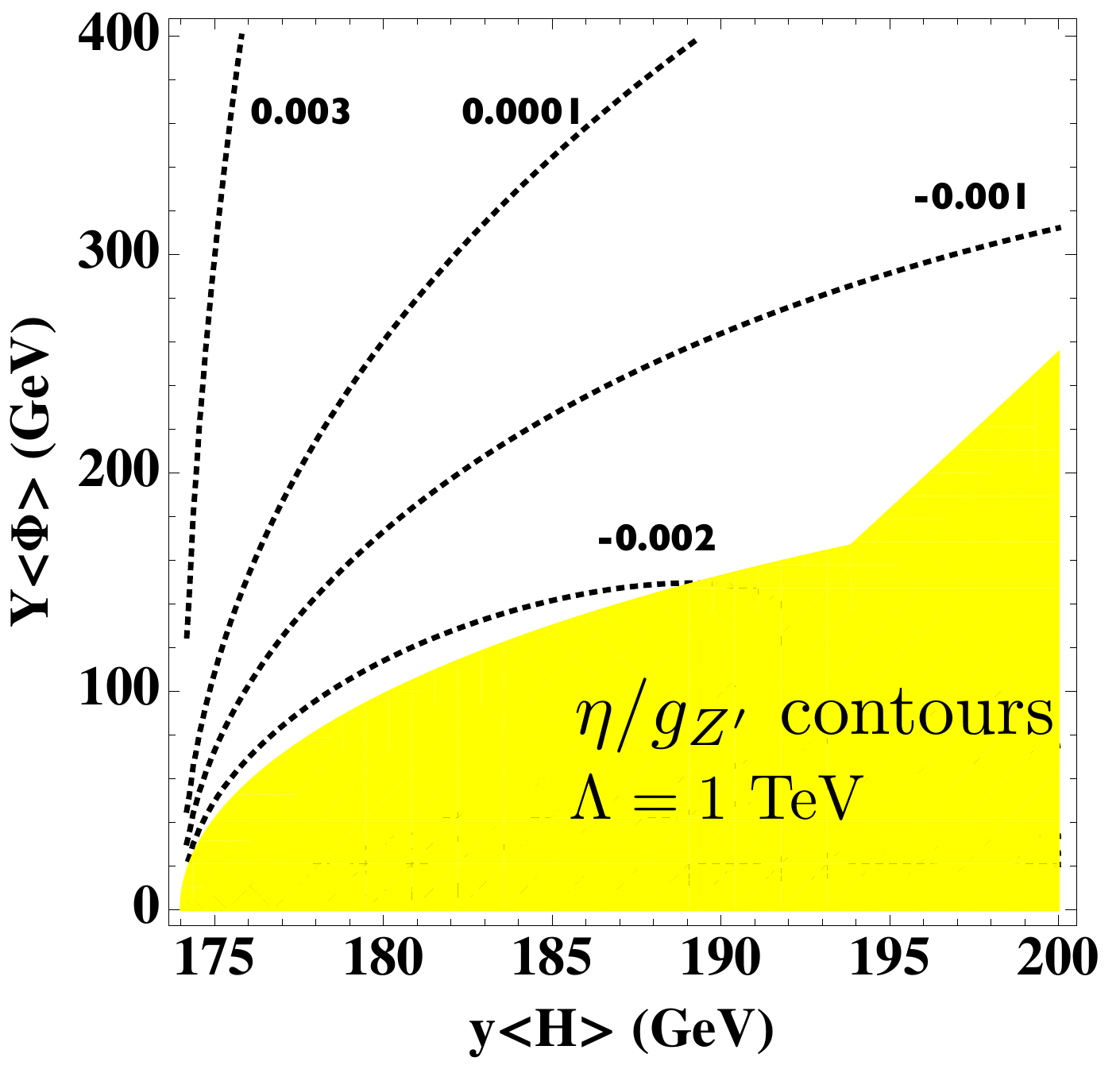}
\end{center}
\caption{ \small Contours for 1-loop contribution to $Z-Z'$ mixing in first (left) and second (right) UV completions presented respectively in Sections \ref{sec:topmodel} and 
\ref{subsec: variant}.}
\label{fig:epsilon} 
\end{figure}

In principle, $\eta$ is a free parameter which is additively renormalized by
loops of the $t$ and $T$ quarks:
\begin{eqnarray}
\eta_{\mbox{ \small 1-loop}}= -\frac{4}{3}\frac{N_c}{16 \pi^2} \left[   (v^{Z}_{tt}  v^{Z'}_{tt}   +a^{Z}_{tt}  a^{Z'}_{tt}  + v^{Z}_{tT}  v^{Z'}_{tT}     
+a^{Z}_{tT}  a^{Z'}_{tT} ) \log(\frac{\Lambda^2}{m_t^2}) 
  + (v^{Z}_{TT}  v^{Z'}_{TT}   +a^{Z}_{TT}  a^{Z'}_{TT}  ) \log(\frac{\Lambda^2}{m_T^2})  \right] 
\end{eqnarray}
  If the $U(1)^\prime$ is embedded in a non-Abelian symmetry, it will naturally vanish
at the scale $\Lambda$ at which that symmetry breaks down to $U(1)^\prime$ (which acts as a UV-cut-off on the log divergence
present in the current framework).  Its natural value thus depends on both the
$Z^\prime$ gauge coupling and the symmetry-breaking scale.
The corresponding contours are shown in Fig.~\ref{fig:epsilon}.
For the parameters of interest to us,
we find $\eta / g_{Z^\prime} \lesssim {\cal O}(10^{-3})$, and thus consistency with the direct bounds 
requires at most modest tuning of $\eta$.
%
%
   \begin{figure}[t]
\begin{center}
\includegraphics[width=0.495\textwidth]{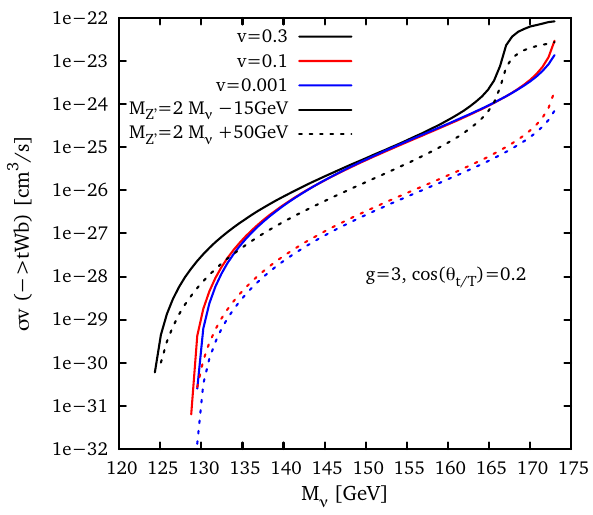}
\caption{\small 
Cross section for annihilation into the $3$-body final state $tWb$ for several
values of the relative velocity $v$ and $M_{Z^\prime}$ mass, as indicated.
We have set $c_L =c_R = 0.2$ for simplicity.}
\label{fig:relic1}
\end{center}
\end{figure}

 \subsection{Flavor Considerations}
 
Finally, we should mention the possibility of flavor violation.  
We have chosen to identify the SM quark with
which the messengers $\psi_{L,R}$ mix as the top quark, but there are no symmetries
that prevent similar mixing terms with the up and charm quarks.  Indeed, the CKM rotations would
 almost certainly induce such terms after electroweak symmetry-breaking.  From a purely
phenomenological stand-point such mixing must be small in order to avoid direct detection constraints as well as constraints from
$D$ meson mixing and rare top decays such as 
$t \rightarrow c Z^{\prime *} \rightarrow c \bar{c} c$.  We will invoke parameters such that
these decays are unobservably small, but they represent a generic feature and an interesting 
signature that is worthy of further investigation.

\section{Dark Matter Annihilation}
\label{sec:annihilation}

\subsection{Relic Density}

In the forbidden channel scenario, the correct relic density can be obtained for DM masses slightly below the mass of the particle running in the loop. For our case this particle is the
top quark, favoring dark matter masses between 150 - 170 GeV.  While such masses will typically have the correct
ballpark relic density, particularly when the 
annihilation occurs somewhat close to the $Z^\prime$ resonance ($\sim 350$ GeV), the detailed story depends
quite sensitively on the entire set of model parameters.
For example, annihilation into
the 3 body final state $tWb$ can be important (see Fig.~\ref{fig:relic1}), 
especially when the dark matter has sufficient
velocity to allow the $Wb$ system to form a close to on-shell top quark.  For regions of
parameters, processes at the loop level can also be significant.  For example, as we will see
below, annihilation into $Zh$ is often the most important channel both today and in the early
Universe.

In addition, there are viable points 
for which the $Z^\prime$ is
only slightly more heavy than the dark matter and the 
relic abundance results from annihilation into $Z^\prime Z^\prime$.  For example,
one has the correct relic abundance for $M_Z^\prime \sim 120$ GeV
when $M_{\nu}=100 $ GeV.
Note that the bound on $\eta$ imposed by direct detection (described in section~\ref{sec:direct})
makes  the contribution from annihilation via $Z$-$Z'$ mixing irrelevant for the relic abundance calculation in virtually
all of the interesting regions of parameter space.

\subsection{One Loop Annihilations and Gamma Ray Lines}

\begin{figure}[t!]
\begin{center}
\includegraphics[angle=0,width=0.99\linewidth]{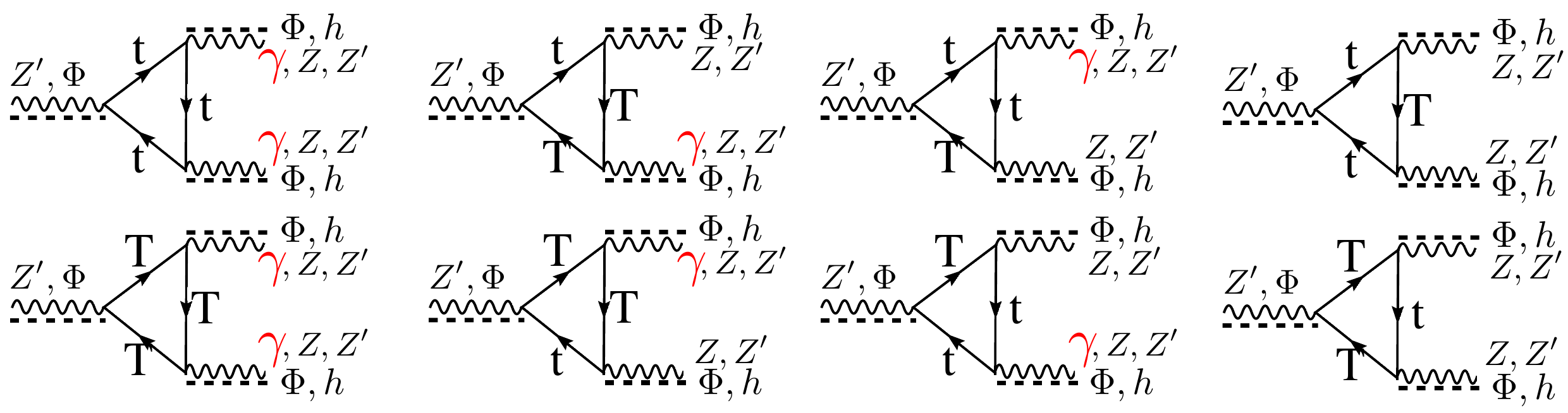}
\end{center}
\caption{One-loop $Z^\prime$ effective vertices induced by $t$ and $T$ couplings  to $\gamma,Z,Z',h,\varphi$.}
\label{fig:effectiVEVertices} 
\end{figure}

\begin{figure}[t!]
\begin{center}
\includegraphics[angle=0,width=0.6\linewidth]{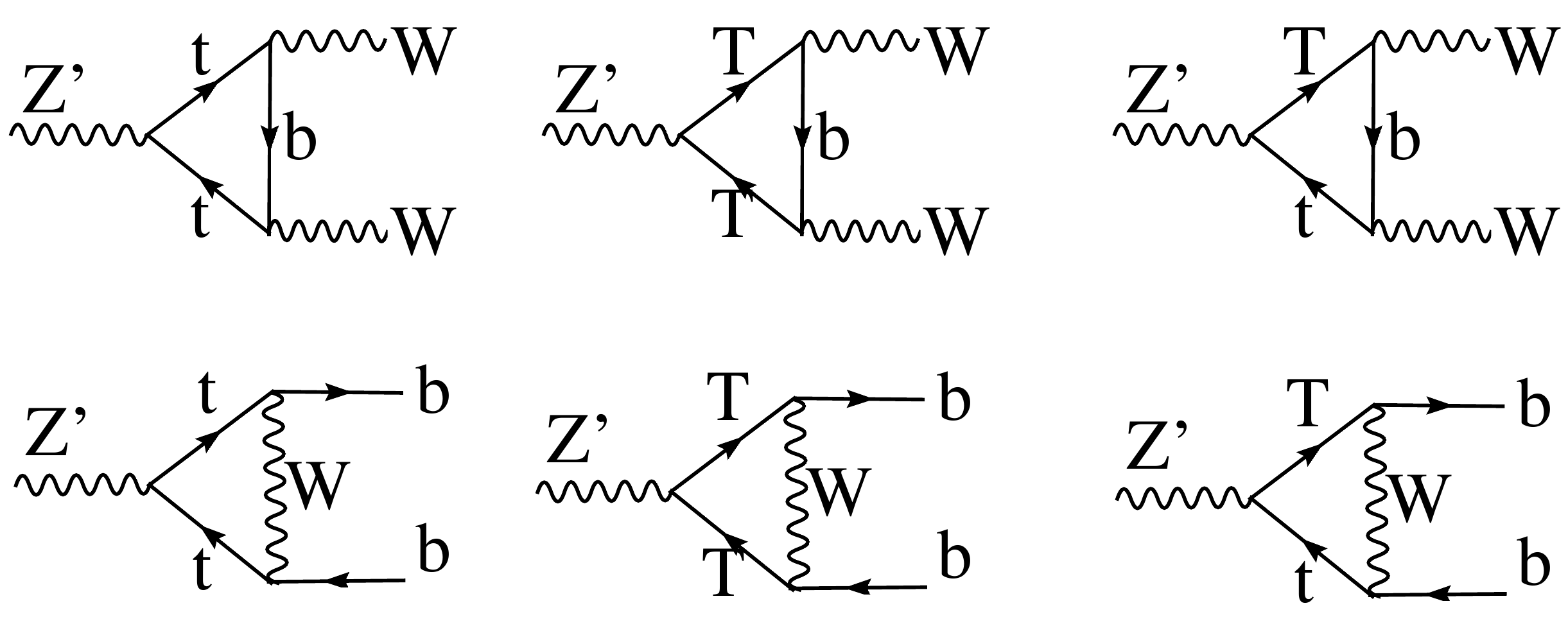}
\end{center}
\caption{One-loop $Z^\prime$ effective vertices induced by $t$ and $T$ couplings to $W$.}
\label{fig:effectiVEVerticesW} 
\end{figure}

At one loop, we have a plethora of final states available as shown in 
Figures~\ref{fig:effectiVEVertices} and ~\ref{fig:effectiVEVerticesW}. 
In addition to line signals from
$\gamma Z$, $\gamma Z'$ and $\gamma h$, there is potentially 
an extra line due to the $\gamma \Phi$ channel. 
Moreover, the one-loop annihilations $Zh$, $Z'h$, $Z \Phi$, 
$WW$, $ZZ$, $ZZ'$ and $b\bar{b}$ can contribute to a sizable continuum of gamma rays.
The analytic results for the effective vertices can be found in the appendices of
Ref.~\cite{Jackson:2013pjq} and (for annihilation into $b \bar{b}$) appendix
\ref{sec:appendixbb}.
Annihilation through the $Z^\prime$-$\Phi$-$h$, $Z^\prime$-$\Phi$-$\Phi$ 
and $Z^\prime$-$h$-$h$ effective vertices vanishes.

Compared to the original Higgs in Space model \cite{Jackson:2009kg},
there are significantly more one loop diagrams.  In that model,
$c_L\ll 1$ and only $t_R$ couples to $Z^\prime$ while the 
$T$-$b$-$W$, $t$-$T$-$Z$, $h$-$T$-$T$ and $\Phi$-$t$-$t$  couplings are suppressed and  
$m_T$ is large.  In the current, UV-complete construction,
$c_L$ and $c_R$ are related and go to zero together. 
In the limit of small mixing  between $\hat{t_R}$ and $\psi$, the $t$-$t$-$Z^\prime$ 
coupling is suppressed and while $T$ has large couplings to both
$Z$ and $Z^\prime$ they are both largely vector-like and thus do not 
contribute~\cite{Jackson:2013pjq}.
However, there is a new feature in the present work that arises because of the 
``mixed loops" where both $t$ and $T$ run in the same diagram. 
The mixed $Z^\prime$-$t$-$T$ couplings are less suppressed for 
small mixing angles and in addition 
there are enhancements due to the presence of two different mass scales in the loop,
leading to a sizable annihilation rate into the vector + a scalar channels, even
for modest mixing angles.

\begin{figure}[t!]
\begin{center}
\includegraphics[angle=0,width=0.9\linewidth]{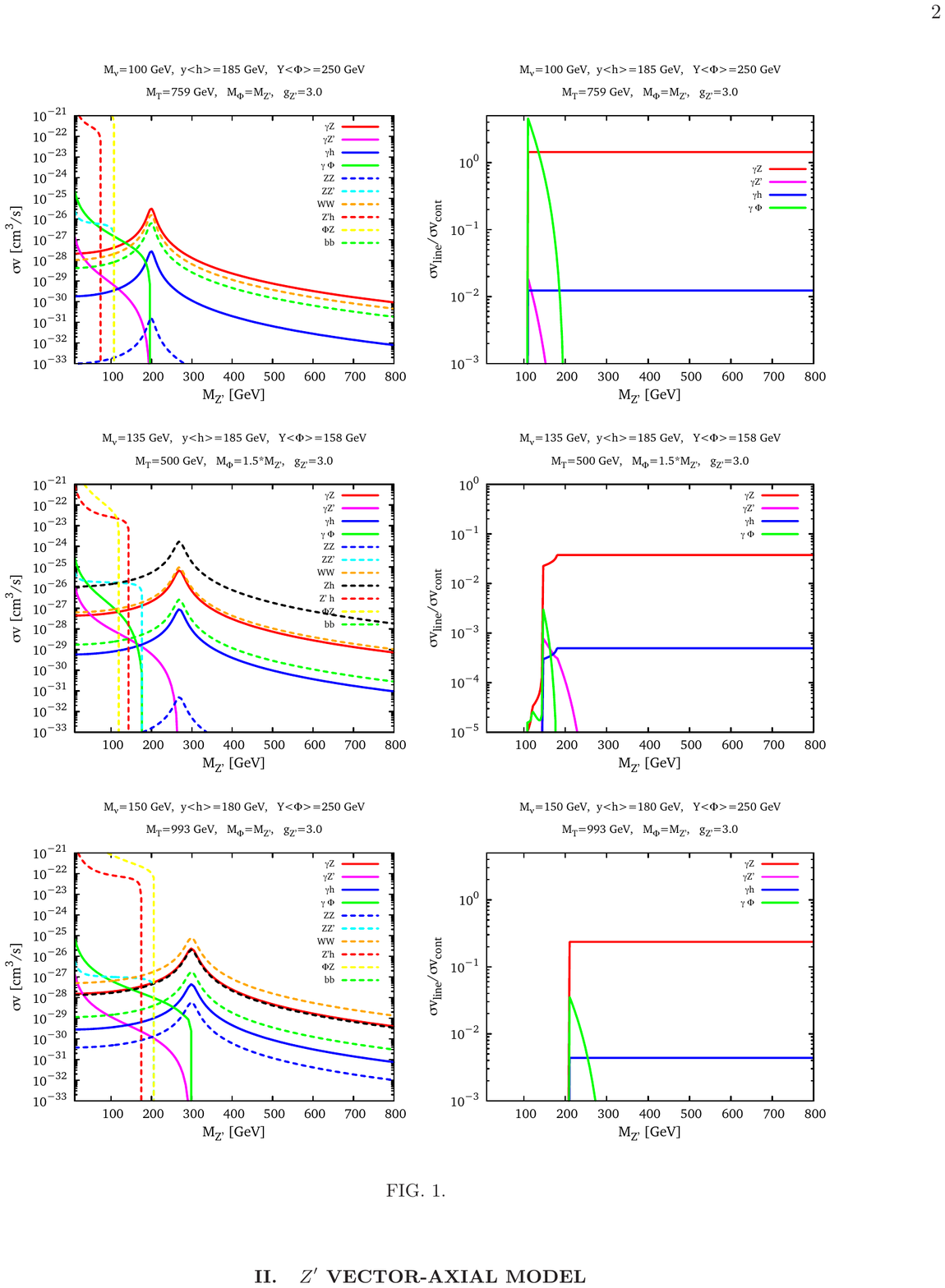}
\end{center}
\caption{ \small Benchmark predictions for cross sections (as labelled in the plot)
and the ratio of line to continuum rates, as a function of the $Z^\prime$ mass. Note that the predictions corresponding to the center panel are essentially unchanged if $M_{\nu}=150 $ GeV rather than  $M_{\nu}=135 $ GeV, for the same choice of $\langle y H \rangle$ and  $\langle Y \Phi \rangle $ values.}
\label{fig:allowedpoint} 
\end{figure}

\begin{figure}[t!]
\begin{center}
\includegraphics[angle=0,width=0.925\linewidth]{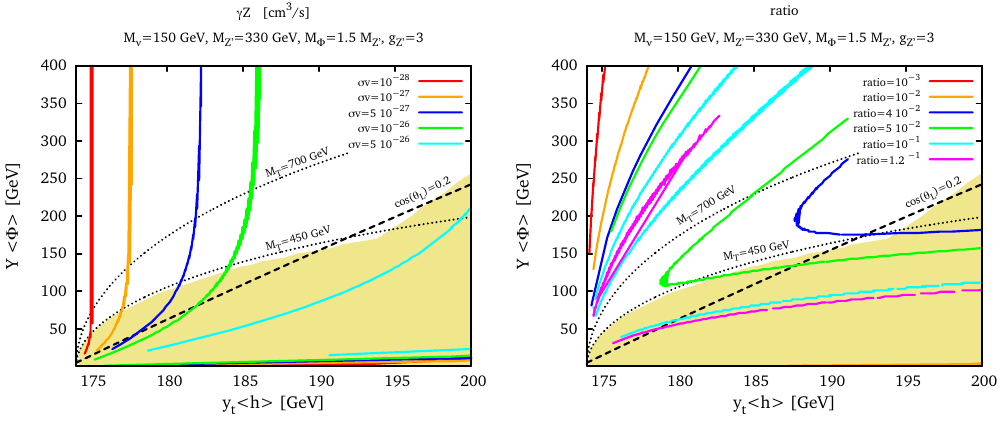}
\includegraphics[angle=0,width=0.925\linewidth]{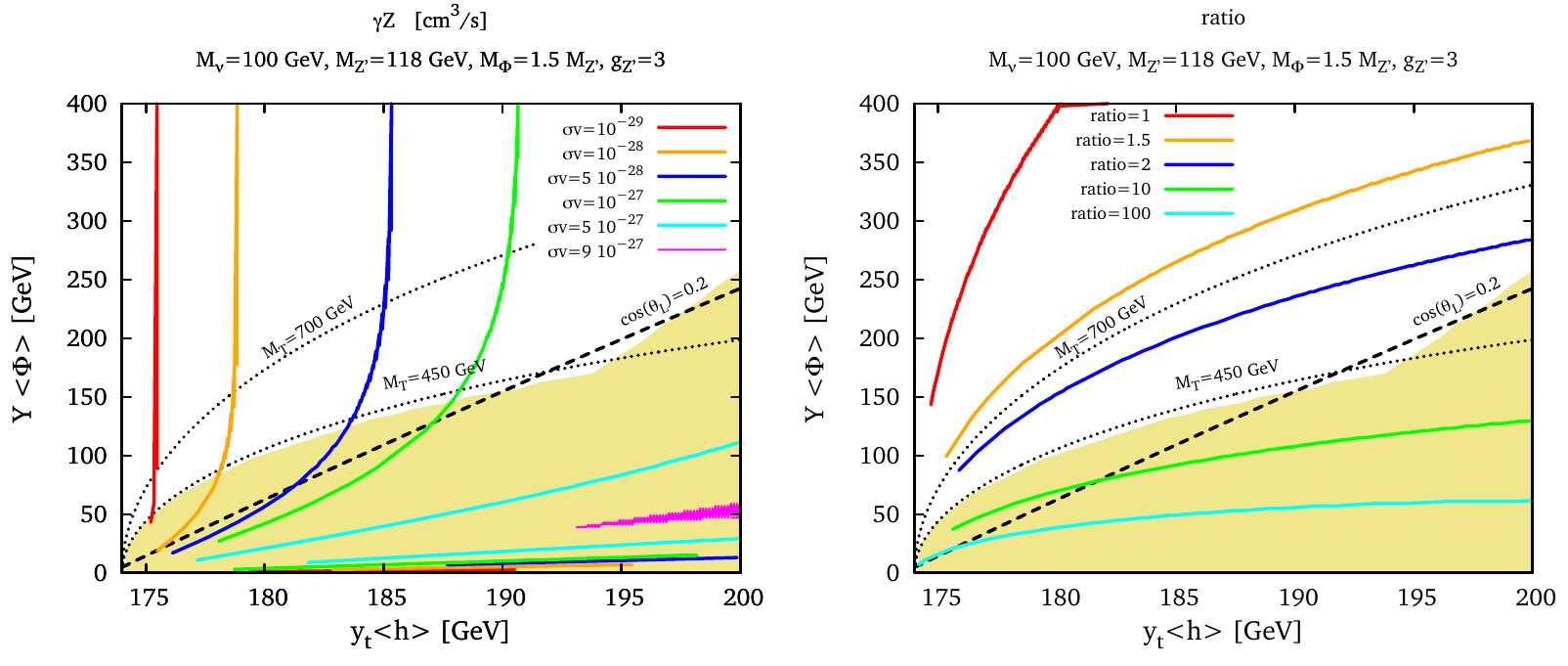}
\end{center}
\caption{ \small Predictions for line signal rates and ratio to continuum for
two choices of DM masses in the plane of ($y \langle H \rangle, Y\langle  \Phi \rangle$).
The shaded region is excluded by EW precision tests. 
In the bottom plots, the choice of the DM and $Z^\prime$ masses guarantees that the correct relic abundance is 
obtained for essentially the whole plane.}
\label{fig:scan} 
\end{figure}

\begin{figure}[h!]
\begin{center}
\includegraphics[angle=0,width=0.45\linewidth]{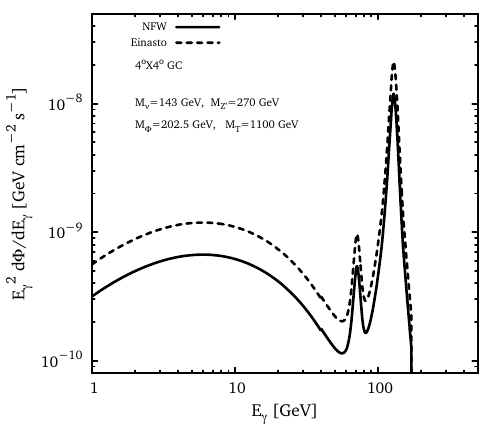}
\end{center}
\caption{ \small Example of a gamma-ray spectrum obtained for 
$\langle y H \rangle =177$ GeV, $\langle Y \Phi \rangle = 200$ GeV, 
$g_{Z'}=g^V_{\nu Z'}=3$, leading to a two-line signal  at $E_{\gamma}=128$ GeV  from $\gamma Z$ annihilation
with  $\sigma_{\gamma Z}v=2 \times 10^{-27}$ cm$^3$s$^{-1}$ and at 
$E_{\gamma}=71$ GeV  from $\gamma \Phi$ annihilation
with  $\sigma_{\gamma \Phi}v=1.35 \times 10^{-28}$ cm$^3$s$^{-1}$. The dominant continuum is from $WW$ and $Zh$ annihilations, $\sigma_{WW}v=5.2 \times 10^{-27}$ cm$^3$s$^{-1}$ and  
$\sigma_{Zh}v=1.7 \times 10^{-27}$ cm$^3$s$^{-1}$.
}
\label{fig:spectrum1} 
\end{figure}

To avoid the overwhelming 1-loop annihilations into gluons 
\cite{Cheung:2010zf,Chu:2012qy,Zhang:2012da}, 
we  focus on the case where DM has vector-like couplings with $Z^\prime$.
Although we are forced to live in a regime of relatively small mixing angles by the EW precision 
constraints,  we can still have large annihilation cross sections into gamma ray lines, as shown 
in Figures~\ref{fig:allowedpoint},  \ref{fig:scan} and \ref{fig:spectrum1}. 

In Ref.~\cite{Jackson:2013pjq}, where no fermion mixing was involved, we found that the dominant channels were
 $Z \Phi$ and $\gamma \Phi$\footnote{This result can be understood as follows: In minimal anomaly-free constructions realizing axial $Z'$ couplings, 1-loop annihilations into two gauge bosons are suppressed due to a cancellation between 1-loop diagrams. Such cancellation does not occur for the $\gamma \Phi$ and $Z \Phi$ channels, which end up dominating.}. For the top quark mediator model, we now have the possibility to produce the SM Higgs in the final state. Moreover, because of the presence of ``mixed" loops involving both $t$ and $T$,
 we find that $Zh$ dominates over $\gamma h$\footnote{The ``mixed" loops are larger than the top loop, but for the $\gamma h$ channel, they cancel each other exactly as $t$ and $T$ have equal couplings to the photon. Therefore,   the $\gamma h$ cross section is entirely controlled by the $t$ loop.
On the other hand, for $Zh$ annihilation, the `mixed" loops  do not cancel each other, as $g_{tZ}\neq g_{TZ}$, and dominate.}
and the discrepancy grows in the limit of small mixing where $g_{ttZ'} \ll g_{tTZ'}$.
In most of the ($y \langle H \rangle, Y\langle  \Phi \rangle)$ parameter space, $Zh$ is the dominant annihilation channel,
although as shown in Figure \ref{fig:scan}, there are regions of parameter space where the $Zh$ continuum 
is relatively suppressed, resulting in a large $\gamma Z$ over continuum ratio.

Roughly speaking, we can classify the situation as to whether the dark matter is heavier or lighter than $(M_Z + M_h)/2$.
For the heavier case, the dominant one-loop annihilation channel is $Zh$.  As discussed above, the relic
abundance will typically work out for dark matter masses somewhat below the top mass, where there
is significant phase space loading of the $Zh$ annihilation rate. Still,  a reasonably
large line to continuum ratio, as large as ${\cal O} (0.1)$ can result, leading to striking line features
as illustrated in Fig.~\ref{fig:spectrum1}.
In the lighter case, the one-loop continuum tends to be dominated by $WW$ (which is always 
significantly larger than $ZZ$ due to two very different masses $m_t$ vs. $m_b$ in the loop, leading to some amplitude enhancement) which is nonetheless is subdominant to the line annihilation into $\gamma Z$,
resulting in a line to continuum ratio of order one (see Fig.~\ref{fig:allowedpoint}).
The correct relic abundance can be obtained from annihilation into $Z'Z'$ with $M_Z' \sim 120$ GeV if $M_{\nu}=100 $ GeV. 

Finally, note that a line from annihilation into $\gamma Z^\prime$  is possible, although only for a relatively
light $Z^\prime$ which will not typically lead to the correct relic abundance.  In addition, there will typically
be a very large one-loop continuum contributions from $Z^\prime h$ and/or $Z \Phi$.
The $\gamma \Phi$ line may be as intense as the $\gamma Z$ one, 
however, there is typically a relatively small $Z^\prime$ mass window where $\gamma \Phi$ is kinematically open while 
$Z \Phi$ is not.  When $Z \Phi$ is open, it tends to dominate all channels by several orders of magnitude.

To summarize this section, we find that when the top quark acts as the primary mediator between dark matter
and the Standard Model via an $s$-channel resonance, the dominant line signal
tends to be $\gamma Z$ with $\gamma h$ relatively suppressed.  The one loop contributions to the continuum
annihilation cannot be ignored, and can be dangerously large from annihilation into $gg$ if the dark matter has
axial vector interactions, and completely swamp any line signal if $Z \Phi$ is open.
 
\section{An Alternate UV Completion}
\label{subsec: variant}

Large line signals require axial couplings in the loop \cite{Jackson:2013pjq}.  The current construction realizes this
need by mixing with the SM top quark, making use of its chiral nature, but its effectiveness is limited by the
constraints from precision data, which require at most a modest level of mixing.
In this Section, we  consider a simple extension of the basic UV completion presented above 
in which the heavy partner fermions $\psi$ have axial couplings with the
$Z^\prime$ even before mixing.  Consequently, there are 
large line signals even in the ``no-mixing" limit, where the top coupling to the $Z'$ coupling goes to zero
(preventing annihilation into $t \bar{t}$), and
allowing for dark matter masses well above the top mass.

We start with the previous UV completion of Section~\ref{sec:topmodel}, and include a second vector-like
doublet $\psi_2=(\psi_{2L},\psi_{2R}) = (\psi^+_{L}, \psi^-_{R})$
in addition to the original
$\psi_1=(\psi_{1L},\psi_{1R}) = (\psi^-_{L}, \psi^+_{R})$. 
Both carry the same SM gauge charges as $\hat{t}_R$, and 
$\psi_1$ also carries $U(1)^\prime$ charge $q_1=q_{\Phi}=1$, whereas $\psi_2$ has $q_2=0$.
For simplicity, we set the vector-like masses of 
$\psi_1$ and $\psi_2$ to zero and
assume that the Higgs Yukawa coupling between $Q_{3L}$ and $\psi_{2R}$ is negligible. We
have Yukawa interactions:
\be
{\cal L} =y \ H \overline{\hat{Q}}_3 \hat{t}_R  + Y \  \Phi \overline{\psi^-_{L}}\hat{t}_R  +\lambda \  \Phi \overline{\psi^-_{L}}{\psi^-_{R}} +\lambda' \ \Phi  \overline{\psi^+_R}{\psi^+_R}
\ee

We recycle the formulae Eq.~(\ref{eq:matrix}--\ref{eq:tan}) for the mass eigenstates $t$ and $T$, with the
replacement that the $\mu$ parameter is replaced by $\lambda\langle \Phi \rangle$ in the formula for $\tan \theta_{R/L}$.
Imposing that the lightest mass eigenstate has the SM top mass fixes the value of 
$\mu \equiv \lambda \langle \Phi \rangle$.  In addition to $t$ and $T$, there is another
massive state $T' \equiv \psi^+$ that does not mix with $\hat{t}_R$ and which has mass given by 
$M_{T'}=\lambda' \langle \Phi \rangle$.  In this way, we realize the existence of
two massive states $T$ and $T'$ with axial-vector couplings to the $Z'$, 
even in the limit  that the SM top quark is purely unmixed.

With respect to the previous case, the couplings to the $Z^\prime$ and $\Phi$ are modified, whereas the couplings to the 
$Z$ and SM Higgs are not. In particular, $t_R$ no longer couples to the $Z^\prime$.
The modified  couplings can be summarized:
\begin{eqnarray}
g^V_{ttZ'}&=& \frac{g_{Z'}}{2} c_L^2,     \ \ \ \  \ \ \ \ g^A_{ttZ'}= -\frac{g_{Z'}}{2} c_L^2,      \ \ \ \  \ \ \ \ 
g^V_{TTZ'}= \frac{g_{Z'}}{2} s_L^2,    \ \ \  \ \ \ \  \    g^A_{TTZ'}= -\frac{g_{Z'}}{2} s_L^2,\\
g^V_{tTZ'}&=& \frac{g_{Z'}}{2} c_L s_L,  \ \  \ \ \ \  g^A_{tTZ'}= -\frac{g_{Z'}}{2} c_L s_L,\\
y^S_{tt\Phi}&=& - Y c_L s_R + \lambda c_L c_R,    \ \ \ \
y^S_{TT\Phi}= Y s_L c_R + \lambda s_L s_R,  \ \  \ \    y^P_{tt\Phi}= 0,  \ \ \ \ y^P_{TT\Phi}= 0\\
\nonumber
y^S_{tT\Phi}&=&  \frac{1}{2}[Y (c_L c_R-s_Ls_R) + \lambda (c_L s_R+s_L c_R)], \ \ 
y^P_{tT\Phi}=  \frac{1}{2}[Y (c_L c_R+s_Ls_R) + \lambda (c_L s_R-s_L c_R)]\\
\end{eqnarray}
and in addition
\begin{eqnarray}
g^V_{T'T'Z'}&=& \frac{g_{Z'}}{2},  \ \  \ \ \ \  g^A_{T'T'Z'}= \frac{g_{Z'}}{2},\\
y^S_{T'T'\Phi}&=&  \lambda' ,  \ \  \ \    y^P_{T'T'\Phi}= 0
\end{eqnarray}

To simplify the parameter space, we set all pseudo scalar couplings of $\psi$ to $\Phi$ in the Lagrangian to zero,
although the mass eigenstates end up with pseudo-scalar couplings for the $\Phi$-$t$-$T$ vertex.
Having set the DM pseudo-scalar couplings to zero, the one-loop diagrams mediated by $\Phi$ 
will be $p$-wave suppressed, leaving the relevant contributions from $Z^\prime$ exchange.

\begin{figure}[t!]
\begin{center}
\includegraphics[angle=0,width=0.85\linewidth]{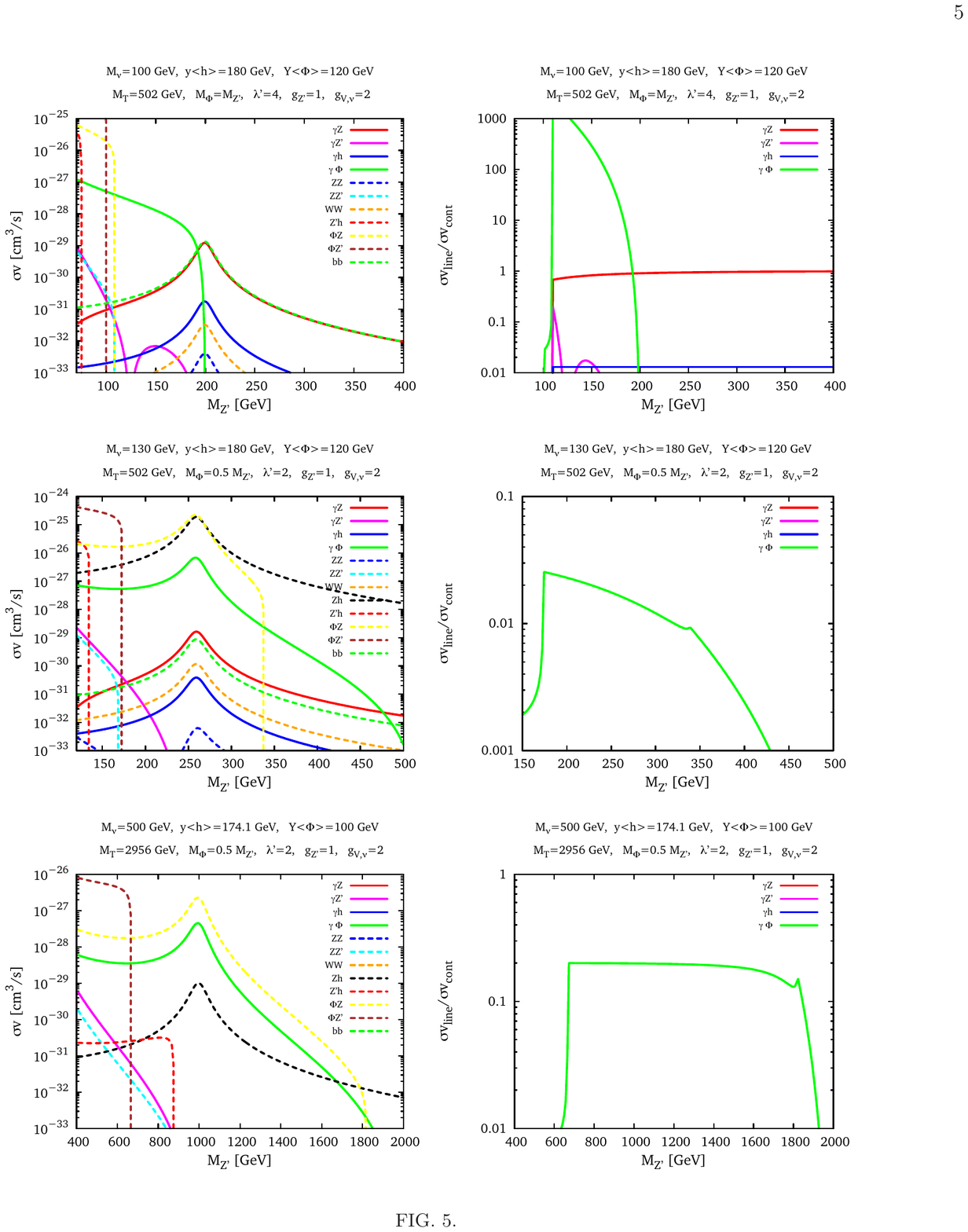}
\end{center}
\caption{ \small Predictions for cross sections and line to continuum ratios in the second UV completion. 
The upper plot corresponds to a light DM case where the $\gamma \Phi$ line signal is huge and 
the continuum is essentially absent for $Z'$ masses leading to the correct relic abundance.
The center plots are quite typical of what happens for a generic choice of parameter, with $Zh$ and $Z \Phi$ channels dominating. The lower plots correspond to the case where $\langle y H \rangle \to m_t$ and mixing angles vanish. In this situation $Z \Phi$ and $\gamma \Phi$ are the main channels~\cite{Jackson:2013pjq}.}
\label{fig:examplesUV2} 
\end{figure}

%
\begin{figure}[t!]
\begin{center}
\includegraphics[angle=0,width=1.\linewidth]{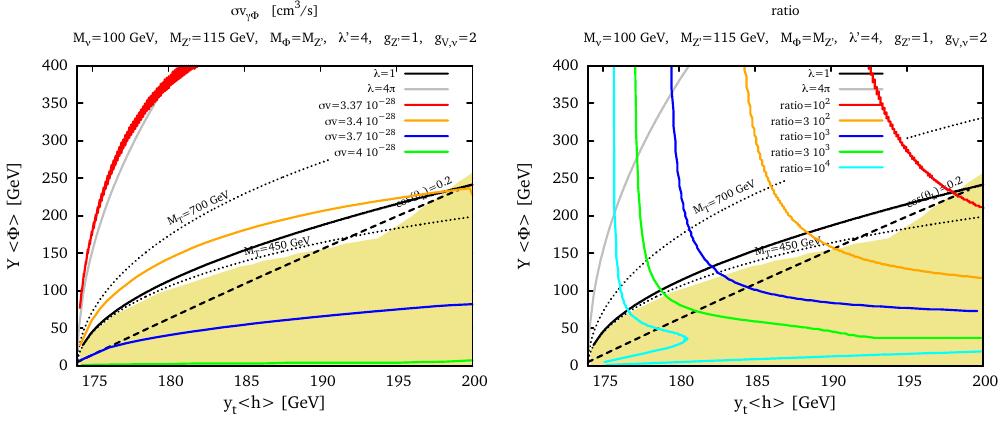}
\includegraphics[angle=0,width=1.\linewidth]{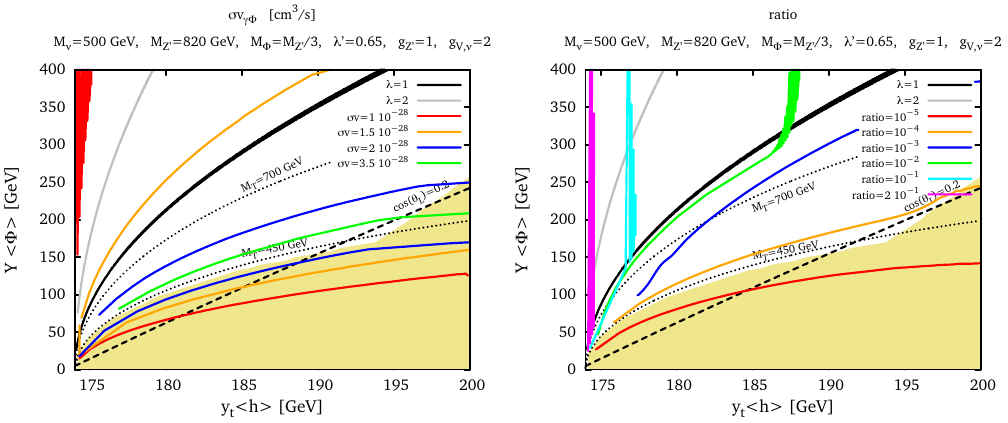}
\end{center}
\caption{ \small  Predictions in the second UV completion, scanning over 
the plane of $y \langle h \rangle$-$Y \langle \Phi \rangle$,
where the shaded region is excluded by EW precision tests. 
In the upper plots, the choice of the DM and $Z'$ masses guarantees that the correct relic abundance is recovered in the whole 
plane through $\nu\bar{\nu} \to Z'Z'$. 
For the lower plots, the relic abundance is controlled by annihilation into $T'T'$
 ($M_{T'}\sim 530$ GeV), $Z' \Phi$ or $tT$ in the region $M_T \lesssim 800$ GeV, in the early Universe,
with the correct relic abundance recovered in a significant fraction of the $(\langle y H \rangle, \langle Y \Phi \rangle)$ plane, 
(particularly  in the region where the line over continuum ratio is larger than $10^{-2}$).}
\label{fig:scanUV2} 
\end{figure}
%
%
%
\begin{figure}[h!]
\begin{center}
\includegraphics[angle=0,width=0.45\linewidth]{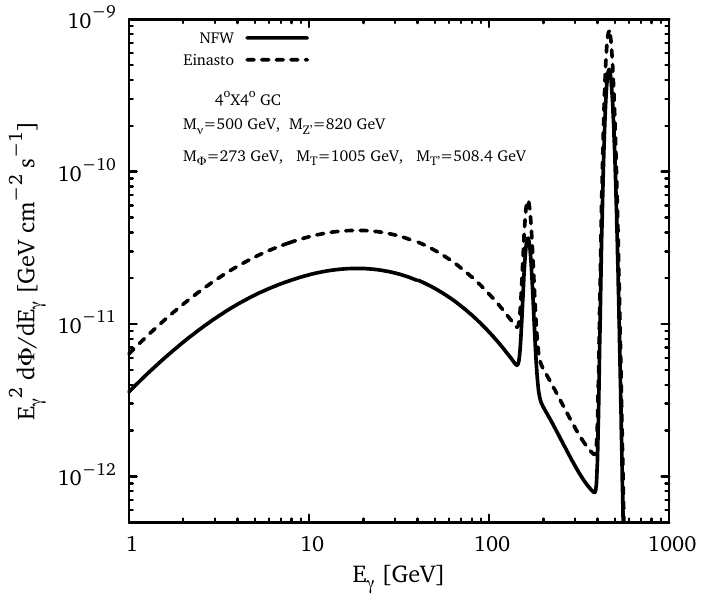}
\end{center}
\caption{ \small Example of a gamma-ray spectrum obtained in the second UV completion for 
$\langle y H \rangle =174.6$ GeV, $ Y = 0.1$, $\lambda'=0.62$,
$g_{Z'}=1=g^V_{\nu Z'}/2$, leading to a two-line signal  at $E_{\gamma}=164$ GeV  from $\gamma Z'$ annihilation
with  $\sigma_{\gamma Z'}v=4 \times 10^{-29}$ cm$^3$s$^{-1}$ and at $E_{\gamma}=463$ GeV  from $\gamma \Phi$ annihilation
with  $\sigma_{\gamma \Phi}v=2 \times 10^{-28}$ cm$^3$s$^{-1}$. The dominant continuum is from $Z\Phi$ annihilations, $\sigma_{Z\Phi}v=5 \times 10^{-28}$ cm$^3$s$^{-1}$. Relic abundance is controlled by $\nu \bar{\nu} \to T' T'$ which is open in the early universe but not today.}
\label{fig:spectrum2} 
\end{figure}
%

As mentioned above, this framework evades large annihilation into $t \bar{t}$ when the dark matter mass is above
the top mass.  However, because the constraints from precision measurements are greatly relaxed, 
$T$ and $T'$ can as light as the direct search bounds of $\sim 475$~GeV,
and one can obtain the correct relic abundance from annihilation into $t\bar{T} / T \bar{t}$, 
$T\bar{T}$, $T'\bar{T}'$ or $Z' \Phi$.
One-loop DM annihilations are dominated by the contributions of $T$ and $T'$,
although mixed ($t,T$) loops still play an important role.
$Zh$ dominates the continuum when it is kinematically accessible.

There are line signals
from $\gamma Z$, $\gamma Z'$ and $\gamma \Phi$, whose relative sizes depend on the choice of parameters, 
in particular the relative masses of ${\nu}$, ${Z'}$ and ${\Phi}$.
We illustrate some cross section predictions in Figs.~\ref{fig:examplesUV2} and \ref{fig:scanUV2}. 
Typically, we find that the dominant line is due to $\gamma \Phi$. 
In the scan plots of Fig.~\ref{fig:scanUV2}, we therefore show contours for $\sigma_{\gamma \Phi} v$ 
and the ratio  $\sigma_{\gamma \Phi} v$ over continuum, finding that
the value of  $\sigma_{\gamma \Phi} v$ is essentially constant in the plane $\langle y H \rangle, \langle Y \Phi \rangle$, 
as it is nearly independent of the mixing angles. However, the other channels depend very sensitively on them. 
As a result, the line over continuum ratios range over several orders of magnitude depending on where in the plane
one considers.
In Fig.~\ref{fig:spectrum2}, we show an illustrative 2-peak spectrum due to $M_{\nu}=500$ GeV DM 
annihilation into $\gamma \Phi$ and $\gamma Z'$.

\section{Summary}
\label{sec:conclusions}

We have considered a class of models where the dark matter is a Dirac fermion 
whose connection with the Standard Model is primarily through the Top quark.  We considered two variations of
UV complete models, and found that both lead to large gamma-ray line signatures from annihilation into either
$\gamma Z$ or $\gamma \Phi$, where $\Phi$ is the scalar responsible for the breaking of a 
new $U(1)^\prime$ gauge symmetry.  In the first variant, the $Z^\prime$ has sizable coupling to the top quark
and a mass only slightly below it,
realizing the forbidden channel mechanism by which the dark matter annihilates primarily into top quarks in the
early universe, but this annihilation is shut off for cold dark matter particles today.
In the second UV completion, the $Z^\prime$ coupling to top quarks is very suppressed, and the dark matter
mass can be much larger than the top mass.
In both cases, $Zh$ (and, depending on the $\Phi$ mass, $Z \Phi$)
is typically the dominant annihilation channel, illustrating the need to carefully consider one-loop contributions to the
continuum photons in theories with enhanced line signals.

Gamma ray line signatures are an important item on the menu of indirect searches for dark matter.  Models 
such as the ones considered here, which
suppress the continuum from annihilations today, are particularly well-probed by these searches.  What is perhaps more
surprising is the fact that these models are surprisingly immune to constraints from the LHC and direct detection
as well, which would suggest that the gamma ray sky hides unique probes of interesting dark theories.
 
\section*{Acknowledgments}

G. Servant is supported by the ERC starting grant Cosmo@LHC (204072).
G. Shaughnessy is supported by the U. S. Department of Energy under the contract 
DE-FG-02-95ER40896.  
T. Tait acknowledges the
hospitality of the SLAC theory group, and is supported in part by NSF
grant PHY-0970171. He also thanks P. Fox, J. Kearny, and A. Pierce for early 
collaboration and discussions, and
the Aspen Center for Physics, under NSF Grant No. 1066293, where part of this work 
was completed. 
The work of M. Taoso is supported by the European Research Council (ERC)
under the EU Seventh Framework Programme (FP7/2007-2013) / ERC
Starting Grant (agreement n. 278234 - `NewDark' project).

\appendix

%
%
\section{One-loop annihilation into $b\bar{b}$}
\label{sec:appendixbb}

\begin{figure}[t]
\begin{center}
\includegraphics[scale=0.5,bb = 115 454 493 715]{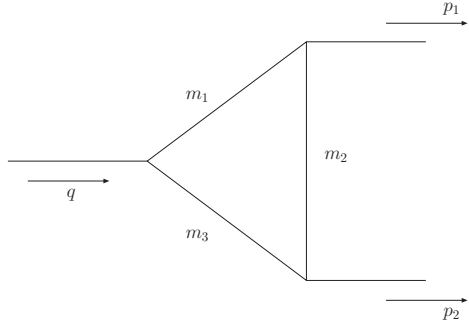} 
\caption[]{Topology for the effective vertices.}
\label{fig:1-to-2}
\end{center}
\end{figure}

In this appendix, we summarize the one-loop expressions for the effective vertex 
and amplitudes-squared needed  for the $b\bar{b}$ channel. One-loop expressions for all other channels are listed in the appendix 
of Ref.~\cite{Jackson:2013pjq}.
The topology for the loop diagrams considered here is shown in Fig.~\ref{fig:1-to-2}.  We express the amplitude in terms of two-point ($B_0$) and three-point ($C_0$) scalar integrals where:
\begin{eqnarray}
C_0 &=& C_0(M_1^2, M_2^2, 4 M_\nu^2; m_1^2, m_2^2, m_3^2) \, , \\
B_0(23) &=& B_0(M_2^2; m_2^2, m_3^2) \, ,\\
B_0(13) &=& B_0(4 M_\nu^2; m_1^2, m_3^2) \, ,\\
B_0(12) &=& B_0(M_1^2; m_1^2, m_2^2) \, .
\end{eqnarray}
In the following expressions, $m_1 = m_3 \equiv m_f$ and $m_2 = M_W$.  However, to derive our results, we have computed all of the expressions in these appendices in terms of the general masses as depicted in Fig.~\ref{fig:1-to-2}.

The effective vertex for $Z^{\prime \alpha} \to b\bar{b}$ is given by the expression:
\begin{equation}
{\cal{V}}_{Z^\prime \to b\bar{b}}^{\alpha} = A \, \bar{u}(p_1) \gamma^\alpha (1 - \gamma_5) v(p_2) \,,
\label{eq:Zpbb-vertex}
\end{equation}
where the loop coefficient $A$ is:
\begin{eqnarray}
A &=& \frac{\left( g_{f\bar{b}}^W \right)^2}{32\pi^2 M_\nu^2 M_W^2 } \biggl(8 A_0(m_f) M_\nu^2
   (a_{f\bar{f}}^{Z^\prime}-v_{f\bar{f}}^{Z^\prime})-m_f^4 \biggl((a_{f\bar{f}}^{Z^\prime}+v_{f\bar{f}}^{Z^\prime})
   (B_0(12)+B_0(13) \nonumber\\
&&-2 B_0(23))+8 C_0 M_\nu^2
   (v_{f\bar{f}}^{Z^\prime}-a_{f\bar{f}}^{Z^\prime})\biggr)-2 M_W^4
   \biggl((a_{f\bar{f}}^{Z^\prime}-v_{f\bar{f}}^{Z^\prime}) \biggl(B_0(12)+B_0(13) \nonumber\\
&&-2
   \biggl(B_0(23)+8 C_0 M_\nu^2\biggr)\biggr)+C_0
   m_f^2 (5 a_{f\bar{f}}^{Z^\prime}-3 v_{f\bar{f}}^{Z^\prime})\biggr)+M_W^2
   \biggl(m_f^2 (3 a_{f\bar{f}}^{Z^\prime}-v_{f\bar{f}}^{Z^\prime})
   \biggl(B_0(12) \nonumber\\
&&+B_0(13)-2 \biggl(B_0(23)+8 C_0
   M_\nu^2\biggr)\biggr)+8 M_\nu^2 (a_{f\bar{f}}^{Z^\prime}-v_{f\bar{f}}^{Z^\prime}) \biggl(-2
   B_0(12)-B_0(13)+3 B_0(23) \nonumber\\
&&+8 C_0 M_\nu^2+1\biggr)+8
   C_0 a_{f\bar{f}}^{Z^\prime} m_f^4\biggr)+4 M_\nu^2 m_f^2 (a_{f\bar{f}}^{Z^\prime}
   (2 B_0(12)+B_0(23)-1)+v_{f\bar{f}}^{Z^\prime} (-2
   B_0(12) \nonumber\\
&&+B_0(23)-1))-2 C_0 m_f^6
   (a_{f\bar{f}}^{Z^\prime}+v_{f\bar{f}}^{Z^\prime})+4 C_0 M_W^6
   (a_{f\bar{f}}^{Z^\prime}-v_{f\bar{f}}^{Z^\prime})\biggr) \,.
\end{eqnarray}
The matrix-element-squared takes the form:
\begin{equation}
\left| {\cal M} \right|^2 = \frac{N_c}{4} \frac{M_\nu^4}{\left| \Sigma_{Z^\prime} \right|^2} \left(\frac{\left( a_{\nu\bar{\nu}}^{Z^\prime} \right)^2 m_b^2
   \left(M_{Z^\prime}^2-4 M_\nu^2\right)^2}{4 M_\nu^2
   M_{Z^\prime}^4}+\left( v_{\nu\bar{\nu}}^{Z^\prime} \right)^2\right)
\end{equation}

\section{$t \bar{t}$ production mediated by $Z'$}
\label{sec:appendixttbar}

\begin{figure}[t]
\begin{center}
\includegraphics[width=0.45\textwidth]{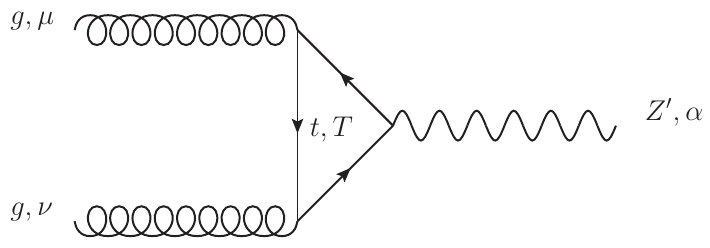}
\caption{ Effective $Z^\prime g g$ vertex mediated through $t$ and $T$ loops. }
\label{fig:ggZp_vertex}
\end{center}
\end{figure}

In this appendix, we provide the expressions for $Z^\prime$ production at the LHC.  The $Z^\prime$ is produced via loop coupling to gluon pairs through the top sector.  Similar to the $Z^\prime \gamma\gamma$ process, the $Z^\prime g g$ vertex convention is shown in Fig.~\ref{fig:ggZp_vertex} and is given by~\cite{Jackson:2013pjq}
\be
{\cal V}^{\mu\nu\alpha}_{g g Z^\prime}= \Gamma_{g g Z^\prime}^{\mu\nu\alpha} \delta^{ab}
\sum_f {N_c^f g_s^2 a_{f\bar f}\over 4 \pi^2}\left( 1+ 2 m_f^2 C_0(M_{Z^\prime}^2, 0, 0, m_f^2, m_f^2, m_f^2)\right),
\ee
where $N_c^f$ is the color factor of the internal fermion, $g_s$ is the $SU(3)$ coupling, $a_{f\bar f}$ is the axial-vector $Z^\prime-f-f$ coupling.  The tensor structure of the $ggZ^\prime$ vertex is given by
\be
\Gamma_{g g Z^\prime}^{\mu\nu\alpha} =\epsilon^{p_1 \mu \nu \alpha}-\epsilon^{p_2 \mu \nu \alpha} + {p_2^\mu \epsilon^{p_1 p_2 \nu \alpha}- p_1^\nu \epsilon^{p_1 p_2 \mu \alpha} \over 2 M_{Z^\prime}^2}
\ee

The loop function $C_0$ can be reduced in this instance to
\be
C_0(M_{Z^\prime}^2, 0, 0, m_f^2, m_f^2, m_f^2)={1\over 2 M_{Z^\prime}^2} \log^2\left({\beta_f-1 \over \beta_f+1}\right),
\ee
where $\beta_f = \sqrt{1-{4m_f^2\over M_{Z^\prime}^2}}$.  Thus, the effective vertex becomes
\be
{\cal V}^{\mu\nu\alpha}_{g g Z^\prime}= \Gamma_{g g Z^\prime}^{\mu\nu\alpha} \delta^{ab}\sum_f {N_c^f g_s^2 a_{f\bar f}\over 4 \pi^2}\left( 1+  {m_f^2 \over M_{Z^\prime}^2} \log^2\left({\beta_f-1 \over \beta_f+1}\right)\right),
\ee

The Landau-Yang theorem forces a vanishing vertex when any of the external bosons is on-shell.  Therefore, the leading production mechanism of the $Z^\prime$ is in association with a gluon or quark shown in 
Fig.~\ref{fig:ttbarprod}.  The $Z^\prime$ subsequently decays to $t\bar t$ pairs.

The amplitude for $g(q_1) + g(q_2) \to t(p_t) + \bar{t}(p_t^\prime)$ can be expressed as:
\begin{eqnarray}
{\cal M}_{gg \to Z^\prime \to t\bar{t}} = \bar{u}(p_t) \gamma^\beta \left( v_{t\bar{t}}^{Z^\prime} + a_{t\bar{t}}^{Z^\prime} \gamma_5 \right) v(p_t^\prime)
\frac{\left( - g_{\alpha \beta} + \frac{p_{Z^\prime, \alpha} p_{Z^\prime, \beta}}{M_{Z^\prime}^2} \right)}{(q_1 + q_2)^2 - M_{Z^\prime}^2 + i M_{Z^\prime} \Gamma_{Z^\prime}} A \Gamma^{\mu\nu\alpha}_{ggZ^\prime} \epsilon(q_1)_\mu \epsilon(q_2)_\nu
\end{eqnarray}
where we have expressed the $Z^\prime gg$ vertex as ${\cal V}_{Z^\prime gg} = A \Gamma^{\mu\nu\alpha}_{ggZ^\prime}$ and $\epsilon(q_1)_\mu$ and $\epsilon(q_2)_\nu$ are the polarization tensors of the incoming gluons. Note that when we square this amplitude and average over the gluon polarizations $\epsilon(q_1)_\mu$ and $\epsilon(q_2)_\nu$, we must do so in such a way that only the physical (transverse) polarization states contribute to the matrix-element-squared. To do this we adopt the general prescription:
\begin{eqnarray}
\sum_{\lambda_i = 1, 2} \epsilon_\mu (q_i, \lambda_i) \epsilon_{\mu^\prime}(q_i, \lambda_i) = - g_{\mu\nu} + \frac{n_{i\mu}q_{i\mu^\prime} + q_{i\mu}n_{i\mu^\prime}}{n_i \cdot q_i} - \frac{n_i^2 q_{i\mu} q_{j\nu}}{(n_i \cdot q_i)^2} \,,
\end{eqnarray}
where $i = 1,2$ and the arbitrary vectors $n_i$ have to satisfy the relations:
\begin{eqnarray}
n_i^\mu \sum_{\lambda = 1,2} \epsilon_\mu(q_i, \lambda_i) \epsilon^*_{\mu^\prime}(q_i, \lambda_i) = 0
\,\,\, , \,\,\,
n_i^{\mu^\prime} \sum_{\lambda = 1,2} \epsilon_\mu(q_i, \lambda_i) \epsilon^*_{\mu^\prime}(q_i, \lambda_i) = 0
\end{eqnarray}
together with $n_i^2 \ne 0$ and $n_1 \ne n_2$. We choose $n_1 = q_2$ and $n_2 = q_1$ such that:
\begin{eqnarray}
\sum_{\lambda_i = 1, 2} \epsilon_\mu (q_i, \lambda_i) \epsilon_{\mu^\prime}(q_i, \lambda_i) = - g_{\mu\nu} + 2 \frac{q_{1 \mu} q_{2 \mu^\prime} + q_{2 \mu} q_{1 \mu^\prime}}{s} \,,
\end{eqnarray}
where $s = (q_1 + q_2)^2$.

Finally, summing over the final-state top quark polarizations and averaging over the initial state gluon polarizations, we find for the matrix-element-squared:
\begin{eqnarray}
\overline{\sum} | {\cal M} |^2 &=& \frac{\left(a_{t\bar{t}}^{Z^\prime} \right)^2 m_t^2 |A|^2 s} {32 \left((s - M_{Z^\prime}^2)^2 + M_{Z^\prime}^2 \Gamma_{Z^\prime}^2 \right)} \left[ 1 -  \frac{s}{M_{Z^\prime}^2}  \right]^2 \,.
\end{eqnarray}
We see that in the limit where the $Z^\prime$ goes on-shell ($s \to M_{Z^\prime}^2$) the amplitude-squared vanishes in accordance to the Landau-Yang theorem.

\end{document}